\documentclass{WileyChemistry-template}
\title{\raggedright Fluorescence Lifetime Imaging Microscopy Analysis of Isolated Melanosomes}

\author{
\begin{minipage}{\textwidth}
Mykyta Kizilov,\textsuperscript{[a]} Sujeong Jung,\textsuperscript{[a]} Vsevolod Cheburkanov,\textsuperscript{[a]} Vladislav V. Yakovlev\textsuperscript{[a]}*
\end{minipage}
}

\newcommand{\affiliation}{
\begin{itemize}
\item[{*}] Email: yakovlev@tamu.edu
\item[{[a]}] Department of Biomedical Engineering, Texas A\&M University, College Station, TX 77843, USA
\end{itemize}
}

\renewcommand{\abstract}{
Melanosomes are organelles found in a wide variety of tissues throughout the animal kingdom. They contain a variety of biological molecules, but the dominant constituent is the pigment melanin, and many functions ascribed to melanosomes, such as photoprotection, are uniquely enabled by the chemical properties and structures of the melanins they contain. In this report, we used, for the first time, Fluorescence Lifetime Imaging Microscopy (FLIM) to examine fluorescent properties of pigments in melanosomes and evaluate their time evolution upon extended laser irradiation. We discovered a relatively short-lived component in fluorescence emission and revealed significant changes in lifetimes upon irradiation indicating structural photoinduced changes to melanin occurring on a time scale of minutes, with observations extending up to one hour.
}
\newcommand{\keywords}{
Fluorescence Lifetime Imaging Microscopy \textbullet\ 
Melanosomes \textbullet\ 
Laser-Induced Damage \textbullet\ 
Photochemistry\textbullet\ 
Pigmentation Disorders
}

\begin{document}
\maketitle\vspace{-1cm}
\textit{\dedication}
\small{\begin{shaded}
		\noindent\abstract
\end{shaded}}

\begin{figure} [!b]
\begin{minipage}[t]{\columnwidth}{\rule{\columnwidth}{1pt}\footnotesize{\textsf{\affiliation}}}\end{minipage}
\end{figure}


\section*{Introduction}

Understanding molecular interactions within living cells is crucial for interpreting the fundamental mechanisms that underpin most cellular functions \cite{day2005imaging}. In particular, the intricate interplay of these interactions not only governs signal transduction \cite{gomperts2009signal}, gene expression \cite{kang2008molecular}, and protein synthesis \cite{liu2012imaging} but also influences the photophysical behavior of specialized organelles, such as melanosomes, which play a key role in cellular responses to light \cite{house2009tracking, meng2016subcellular}. Investigating these molecular dynamics in a live-cell context provides deeper insights into the regulatory networks that govern cellular behavior, advancing fields such as cellular biology, disease pathology, and therapeutic development \cite{liu2015imaging, 2025CheburkanovBrillouinSPIE, 2025CheburkanovGliaSPIE}.

Melanosomes are specialized organelles responsible for the synthesis, storage, and transport of melanin, a critical pigment for photoprotection in the skin, eyes, and hair of mammals \cite{burke2005mosaicism}. Beyond their role in pigment production, melanosomes exhibit unique photophysical properties that directly affect light absorption and fluorescence. In retinal pigment epithelium (RPE) cells, for example, melanosomes absorb light and neutralize reactive oxygen species to protect photoreceptors from photo-oxidative damage. Melanin exists primarily in two forms—eumelanin (a dark nitrogen-containing polymer) and pheomelanin (a sulfur-containing pigment with a yellowish hue) \cite{prota2012melanins, miltenberger2002molecular}; variations in melanin composition can affect melanosome antioxidant capacity. Although eumelanin-rich granules typically provide robust photoprotection \cite{sarna1992new}, recent studies indicate that melanosomes can degrade when exposed to prolonged or intense light radiation \cite{hill1997melanin}.

Despite their importance, the mechanisms underlying melanosome degradation, especially under laser or light exposure, remain poorly understood \cite{miltenberger2002molecular}. Some studies suggest that melanosomes suffer irreversible photodamage \cite{Alam2022}, leading to structural disintegration and diminished antioxidant function \cite{zareba2006effects}. This degradation is of particular concern in clinical fields such as dermatology and ophthalmology, where lasers are used therapeutically to treat pigment-related disorders. Understanding melanosome responses to laser exposure at a molecular level is therefore crucial for optimizing therapeutic applications and preventing adverse effects.

In most recent studies, fluorescence spectra were evaluated for melanolipofuscin granules isolated from retinal pigment epithelium cells \cite{Dontsov2023understanding}. Those results show that visible light irradiation induces oxidative degradation of melanin, leading to the formation of water‐soluble fluorescent products. Such findings provide further insight into the mechanism of light-induced melanin loss in the RPE, emphasizing the role of superoxide radicals and oxidative stress in the deterioration of melanosomes. This work complements our observations by highlighting that similar oxidative processes may contribute to the structural and photophysical changes observed in melanosomes under prolonged laser exposure.

A variety of experimental techniques have been employed to study the morphology, structure, and chemical changes of melanosomes during light exposure \cite{Jimbow1982Characterization, Lea1976The}. Traditional microscopy methods, such as scanning electron microscopy  (SEM) and transmission electron microscopy (TEM), have provided insights into structural alterations. Spectroscopic methods, including Raman spectroscopy and electron paramagnetic resonance (EPR), reveal chemical changes and redox states. However, these methods can face limitations when capturing dynamic changes in live samples. Melanin’s inherently low fluorescence yield further complicates analysis via standard fluorescence microscopy.

Recent advancements in laser-based therapies have revolutionized the treatment of pigmentation disorders, providing clinicians with precise tools to target melanin-rich tissues. Advanced imaging techniques like FLIM offer unique opportunities to study melanosome behavior in real-time \cite{ruedas2015flim}. Multispectral autofluorescence lifetime imaging, combined with machine learning, has also proven effective in diagnosing pigmented skin lesions \cite{vasanthakumari2022discrimination, vasanthakumari2024pixel, 2025KizilovColonAPS} and identifying precancerous and cancerous oral lesions \cite{duran2021machine}. Nonetheless, deeper understanding of photodamage mechanisms is needed, as laser exposure may lead to melanosome fragmentation, oxidative stress, and localized heating \cite{schmidt2016temperature, yi2018degraded}, particularly under repeated or high-intensity applications.

Moreover, the eumelanin-to-pheomelanin ratio may influence the susceptibility to laser-induced degradation \cite{hu2020heat}. Identifying how structural and molecular changes emerge during laser exposure is essential to design safer therapeutic protocols. FLIM, which measures fluorescence decay times rather than intensity, is especially suited to study such processes in heterogeneous samples \cite{luecking2020capabilities, alghamdi2016ultrastructural}.

Short-lifetime FLIM imaging captures ultrafast decay components often missed by conventional FLIM \cite{saha2011raman, yakovlev2008real, suhling2015fluorescence, 2025CheburkanovHemoglobinSPIE}, a feature relevant to studying early-stage photodamage. By separating overlapping signals from multiple fluorophores or excited states, short-lifetime FLIM reveals subtle molecular interactions in melanosomes.

In this study, we use FLIM to investigate melanosomes' responses to laser exposure, tracking both short- ($<10^{-1}ns$) and long- ($>10^{-1}ns$) lifetime fluorescence components. Our aim is to explore the potential mechanisms of melanosome degradation by evaluating changes in the fluorescence lifetimes over time. Specifically, we address how laser exposure alters the molecular environment, whether distinct degradation stages can be identified from lifetime data, and the potential clinical implications for laser-based therapies.

We employed a custom-built confocal FLIM system with time-correlated single-photon counting (TCSPC) and reconvolution fitting to extract decay constants with high precision \cite{2025BerezinFLIMSPIE}. These measurements enabled us to uncover subtle photophysical changes that occur under laser exposure, offering unprecedented detail on melanosome degradation dynamics.

By correlating lifetime patterns with stages of degradation, this work provides a framework for optimizing laser parameters to minimize tissue damage. Our findings may also inform research on pigment-related diseases such as macular degeneration and melanoma \cite{2025KizilovMelanomaSPIE}, where melanosome behavior is key to developing effective diagnostic and therapeutic strategies.

In summary, we present a detailed FLIM-based investigation into the photophysical behavior of melanosomes under laser exposure, highlighting short-lifetime fluorescence components. Our results reveal a range of degradation mechanisms that underscore the promise of FLIM in advancing research on melanosome biology, dermatology, and biomedical optics \cite{yakovlev2018biochemical}.

Recent advancements in laser-based therapies have revolutionized the treatment of pigmentation disorders, providing clinicians with precise tools to target melanin-rich tissues \cite{Kizilov2025, Shimojo2024, Kauvar2012}. Although techniques such as time-resolved fluorescence spectroscopy and Raman spectroscopy \cite{jhraman, mykraman, Cubeddu1990, 2025HarringtonDUVSPIE} have been employed to study melanosome properties, they exhibit inherent limitations. Time-resolved fluorescence spectroscopy generally lacks the spatial resolution needed to resolve individual melanosomes and their heterogeneous behavior, while Raman spectroscopy, despite its molecular specificity, often requires long acquisition times and is prone to interference from fluorescence background \cite{Cubeddu1990}.

Recent in vivo imaging studies have provided critical insights into the clinical significance of pigment granule alterations in the retinal pigment epithelium. For instance, Meleppat et al. \cite{meleppat2021vivo} demonstrated that multimodal retinal imaging techniques, combining directional back-scattering and short-wavelength fundus autofluorescence, can effectively capture disease-related changes in melanosome and lipofuscin densities, serving as important biomarkers for conditions such as Stargardt disease and age-related macular degeneration. Complementarily, Dontsov and Ostrovsky \cite{dontsov2024retinal} investigated the norms and age-related alterations of RPE pigment granules, elucidating the mechanism of light-induced oxidative degradation that leads to a decline in melanin content. These findings underscore the potential of advanced optical imaging to reveal subtle changes in pigment composition, thereby reinforcing the relevance of our FLIM approach to studying melanosome degradation dynamics.

FLIM uniquely combines high temporal resolution with spatial mapping capabilities, allowing real-time visualization of fluorescence decay at the single-organelle level \cite{Seidenari2013, Arginelli2013}. This dual capability not only captures rapid photophysical changes but also reveals local variations in the microenvironment—features that are critical when studying the dynamic processes of melanosome degradation under laser exposure \cite{Fink2016}. Consequently, FLIM offers a more comprehensive insight into both the kinetic and spatial aspects of melanosome photophysics compared to traditional methods \cite{Seidenari2013, Arginelli2013}.


\section*{Experimental Setup for Fluorescence Lifetime Imaging Microscopy}

\begin{figure*}[h!]
    \centering
    \includegraphics[width=0.85\textwidth]{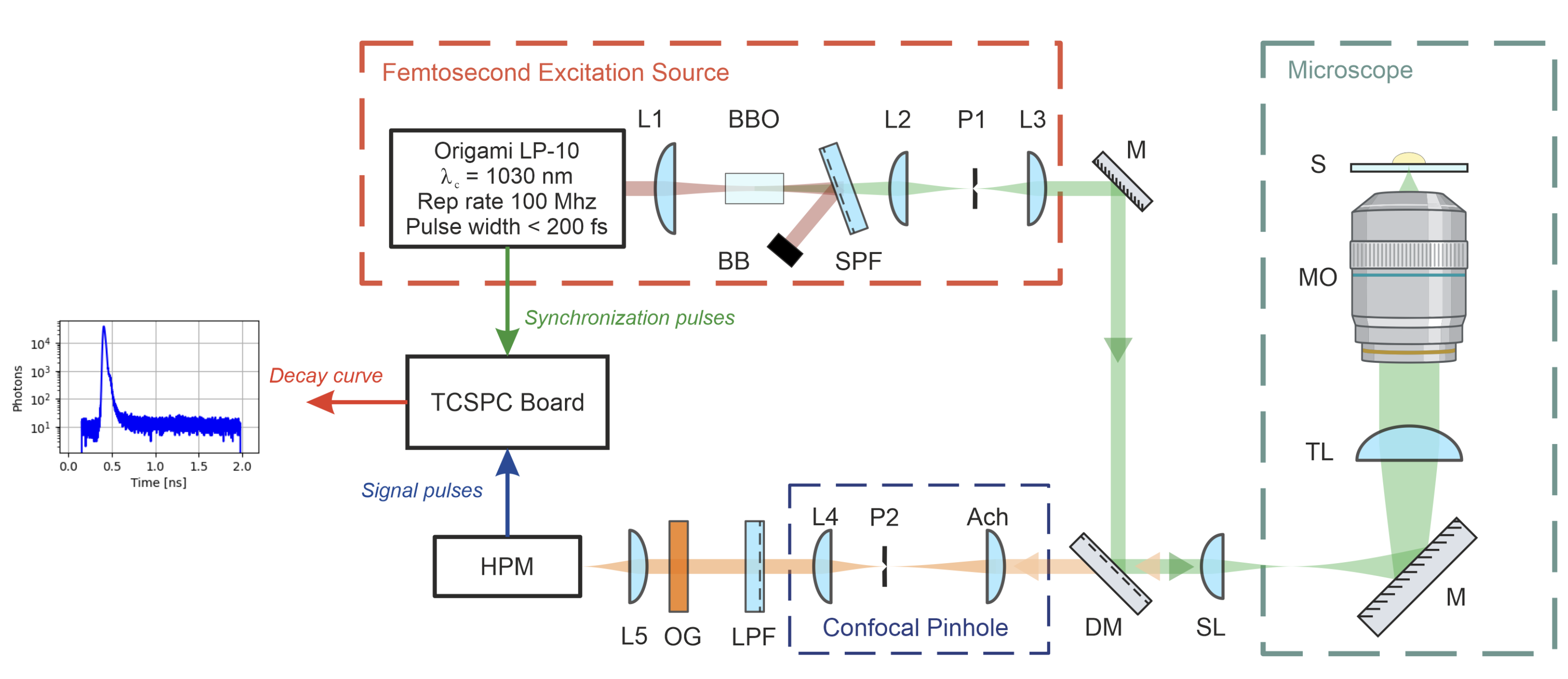}
    \caption{Schematic diagram of the experimental setup for fast FLIM acquisition. The setup includes a femtosecond excitation source, optical components for beam shaping and filtering, a home-built inverted microscope with a confocal pinhole attachment, and a detection system using a hybrid photomultiplier (B$\&$H HPM-100-06). Key components are labeled for clarity. \textbf{BBO} -- Beta-Barium Borate SHG crystal ($\theta=23.4^{\circ}$); \textbf{BB} -- Beam block; \textbf{SPF} -- Short-pass filter; \textbf{L} -- Plano-convex lens; \textbf{P} -- Precision pinhole aperture; \textbf{M} -- Mirror; \textbf{SL} -- Achromatic doublet scan lens, 50 mm focal length; \textbf{TL} -- Achromatic doublet tube lens, 200 mm focal length; \textbf{MO} -- Microscope objective lens; \textbf{S} -- Sample; \textbf{HPM} -- Hybrid photomultiplier; \textbf{OG} -- Orange glass spectral filter; \textbf{LPF} -- Long-pass filter; \textbf{Ach} -- Achromatic Lens; \textbf{DM} -- Dichroic mirror. Figure made with assets created with BioRender.com}
    \label{fig:FLIM_system}
\end{figure*}

A schematic representation of the experimental setup is depicted in Fig. \ref{fig:FLIM_system}.

The output from a femtosecond laser (Onefive Origami LP 10-100) was focused onto a $\beta$-barium borate (BBO, EKSMA Optics) crystal (cut angle $\theta=23.4^{\circ}$) using a plano-convex lens (L1, Thorlabs, 75 mm focal length), generating second harmonic light at a wavelength of 515 nm. With an initial input power of 180 mW at 1030 nm, we achieved a conversion efficiency of 2\%, producing approximately 4 mW of 515 nm light. Melanosomes are characterized by a broadband, almost featureless absorption spectrum which extends to the near-UR and UV parts of the spectrum \cite{cone2015measuring} and a broad fluorescence emission spectrum \cite{furtjes2023intraoperative}. We chose 515 nm, which happens just at the edge of a steep rise of an optical absorption while being off the peak of the fluorescence. 

A hard-coated short-pass filter (SPF, Thorlabs, 600 nm cut-off) was used to remove the pump radiation. The spatial profile of the beam was cleaned by focusing into precision 75 $\mu m$ pinhole aperture (P1, National Aperture) with a plano-convex lens (L2, Thorlabs, 50 mm focal length). The output of the pinhole followed Airy distribution with transmitted power under 2 mW. The beam was subsequently collimated using a plano-convex lens (L3, Thorlabs, 100 mm focal length) and directed to the custom-built microscope via a dichroic mirror (DM, Semrock, long-pass filter with 550 nm cut-off wavelength).

For optimal spatial resolution in the detection path, the excitation beam passed through a fixed angular magnification telescope, comprising an achromatic doublet (SL, Thorlabs, 50 mm focal length) and an achromatic doublet tube lens (TL, Thorlabs, 200 mm focal length), which expanded the beam to overfill the entrance pupil of the micriscope objective lens (MO, Nikon CFI60 series, 20x, 0.5 NA). Fluorescence emission was collected by the same objective used for its excitation. The emitted fluorescence was passed through the DM and directed to the confocal pinhole assembly.

To suppress out-of-focus light, the emitted signal was focused into a pinhole aperture (P2, National Aperture) with an achromatic doublet (Ach, Thorlabs, 40 mm focal length). Signal from the pinhole was collimated by a plano-convex lens (L4, Thorlabs, 35 mm focal length) to minimize distortions caused by the necessary spectral filters. Residual excitation light is filtered by a combination of a long-pass filter (LPF, Semrock, OD $>$ 4.0 at 515 nm) and an orange glass  filter (OG, Thorlabs, OD $>$ 5.0 at 515 nm). Overall optical density (OD) of approximately 7.0 provided sufficient attenuation of unwanted spectral components, given a laser repetition rate of 100 MHz and a maximum signal flux of $10^7$ photons per second. Filtered signal was then focused onto the entrance window of a hybrid photomultiplier (HPM-100-06, Becker$\&$Hickl) with a plano-convex lens (L5, Thorlabs, 75 mm focal length).

Signal from the HPM along with the synchronization pulse train from the laser were sent to the correlator (TCSPC Board) to retrieve the fluorescence decay curve, shown in the \ref{fig:FLIM_system}.

The excitation power was set to $0.35$ mW, ensuring sufficient signal strength while maintaining stable fluorescence lifetime measurements. While laser-induced degradation of melanosomes is an important consideration, optimizing laser power to systematically study its effects on degradation was beyond the scope of this study. Future investigations will explore the dependence of melanosome degradation on laser power.


\section*{Sample Preparation and Experimental Design}

\subsection*{Sample Preparation}
The preparation procedure of bovine and porcine pigment granules (melanosomes) followed the method of Dontsov et al. \cite{dontsov1999retinal, denton2013hyperthermia}. This method provided samples enriched in light and heavy fractions of melanosomes by density for both bovine and porcine samples. Light melanosomes are predominantly spherical, while the heavy melanosomes are predominantly elliptical in shape, resulting in slight size differences between the two fractions \cite{roegener2004pump}. Retinal pigment epithelium pigment granules stimulate the photo-oxidation of unsaturated fatty acids \cite{dontsov1999retinal}. However, no high-resolution imaging was used, and therefore, no precise knowledge of shape variance within each fraction was obtained. The melanosome fractions were separated, and stock solutions of the melanosomes were stored at 4°C. Several dilutions of melanosomes were prepared using deionized water for each fraction prior to analysis. Plated aqueous melanosomes were prepared on glass microscope slides in order to observe cavitation events. A loosely sealed silicone washer and glass cover slip were used to enclose the melanosome sample to prevent evaporation of the aqueous suspension.

TCSPC was employed for data acquisition. Each acquisition session lasted 60 cycles of 1 minute, with fluorescence decay curves recorded over a 2 ns window, producing a channel resolution of approximately 500 fs.

\subsection*{Retrieval of the Instrument Response Function (IRF)}

The system's detectors exhibit an IRF with a pulse duration width at the half maximum (commonly referred to as the Full Width at Half Maximum, FWHM) of approximately 20 ps, allowing for the observation of relaxation phenomena on picosecond timescales. To ensure accurate extraction of fluorescence lifetimes, it is essential to obtain the true IRF for each acquisition rather than relying on a simulated one.

The IRF was measured by removing the OG filter and placing a silver-coated mirror on the sample stage in the experimental setup (Fig. \ref{fig:FLIM_system}. The axial position of the microscope objective was adjusted to maximize the reflection signal detected by the HPM. The detected signal, consisting of the laser pulses reflected from the mirror surface, provided the necessary data to retrieve the system's IRF.
\subsection*{Imaging Protocol}

To prepare samples for imaging, 5 $\mu L$ of melanosome solution was pipetted onto a concave microscope slide (Cole-Parmer) and sealed with a glass coverslip (VWR International) to prevent exposure to air and maintain sample integrity.

Before imaging, the system was aligned using a short-lifetime fluorophore (4-[4-(dimethylamino)styryl]-1-methylpyridinium iodide (4-DASPI, MilliporeSigma) in water), which served as an alignment marker. A 25 $\mu L$ aliquot of 4-DASPI solution was placed on a concave slide and covered with a glass coverslip. The sample was positioned in the focal plane of the microscope objective, and adjustments were made along the optical axis to maximize the detected fluorescence signal.

After alignment, a blank slide and coverslip were inserted to confirm the absence of detectable photons, ensuring any detected signals were exclusively from the sample of interest. This validation confirmed proper alignment and preparation of the optical system for the subsequent experiment.

\begin{figure}[h!]
    \centering
    \includegraphics[width=0.4\columnwidth]{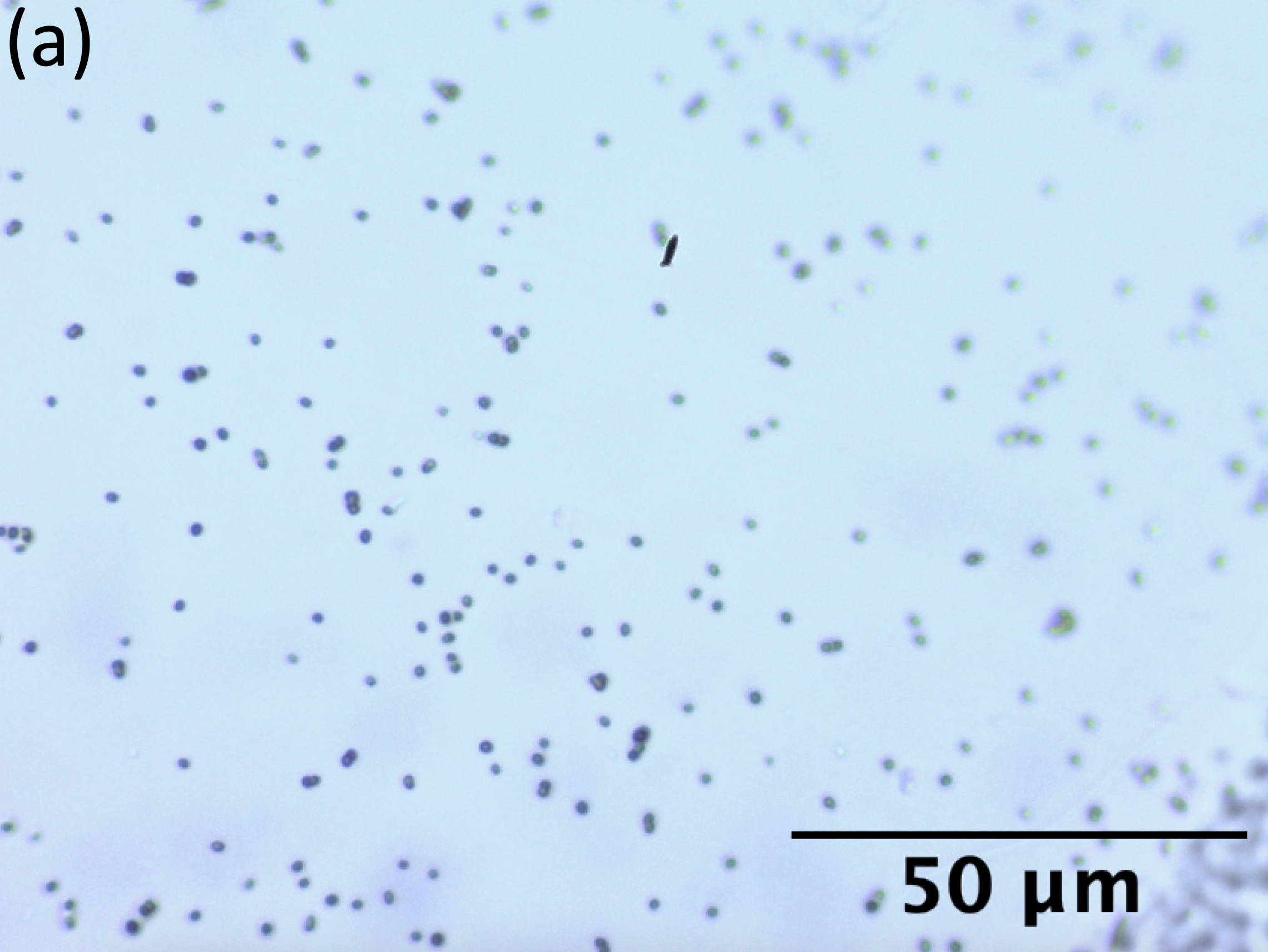}
    \includegraphics[width=0.4\columnwidth]{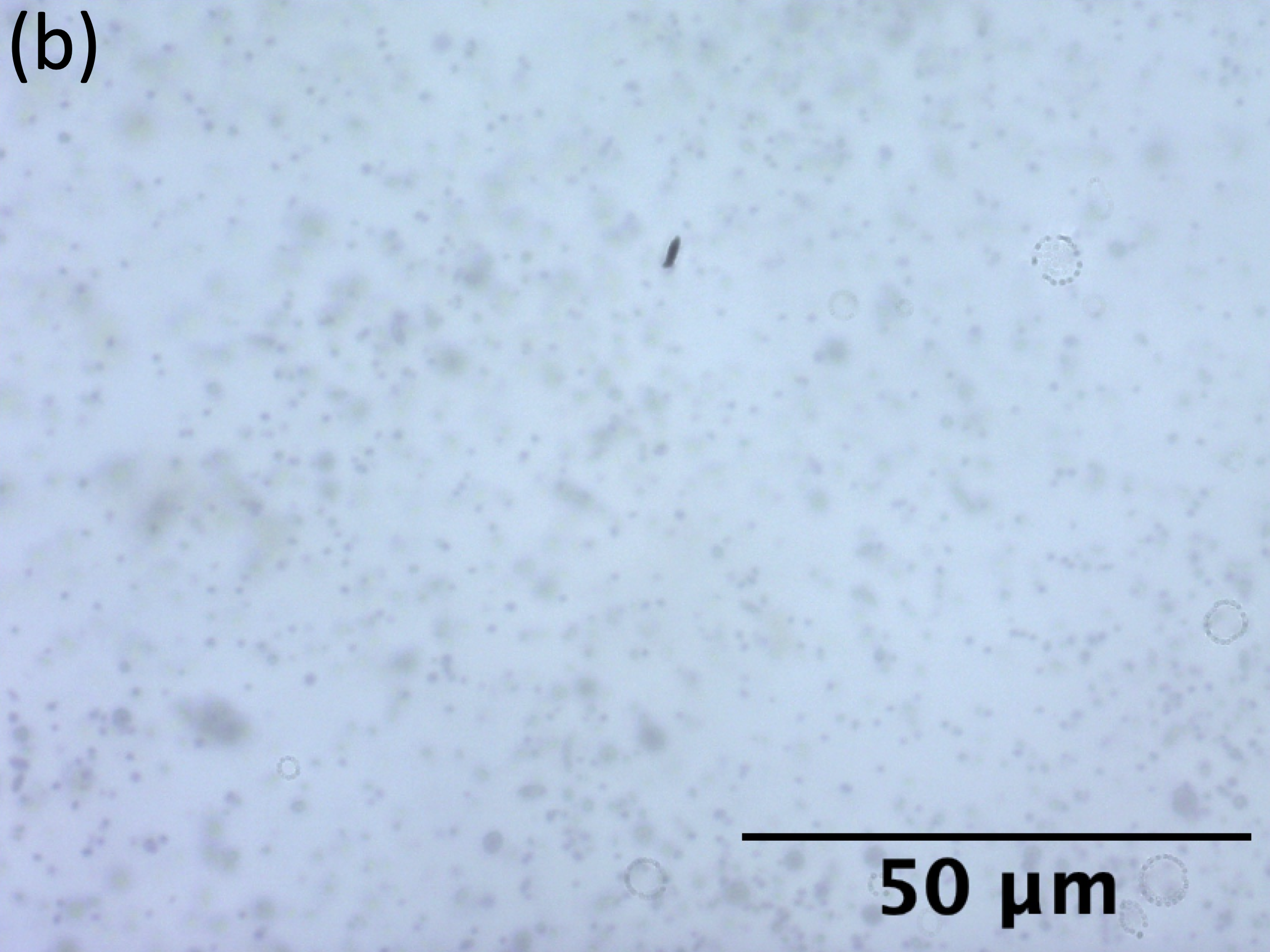}
    \includegraphics[width=0.3\columnwidth]{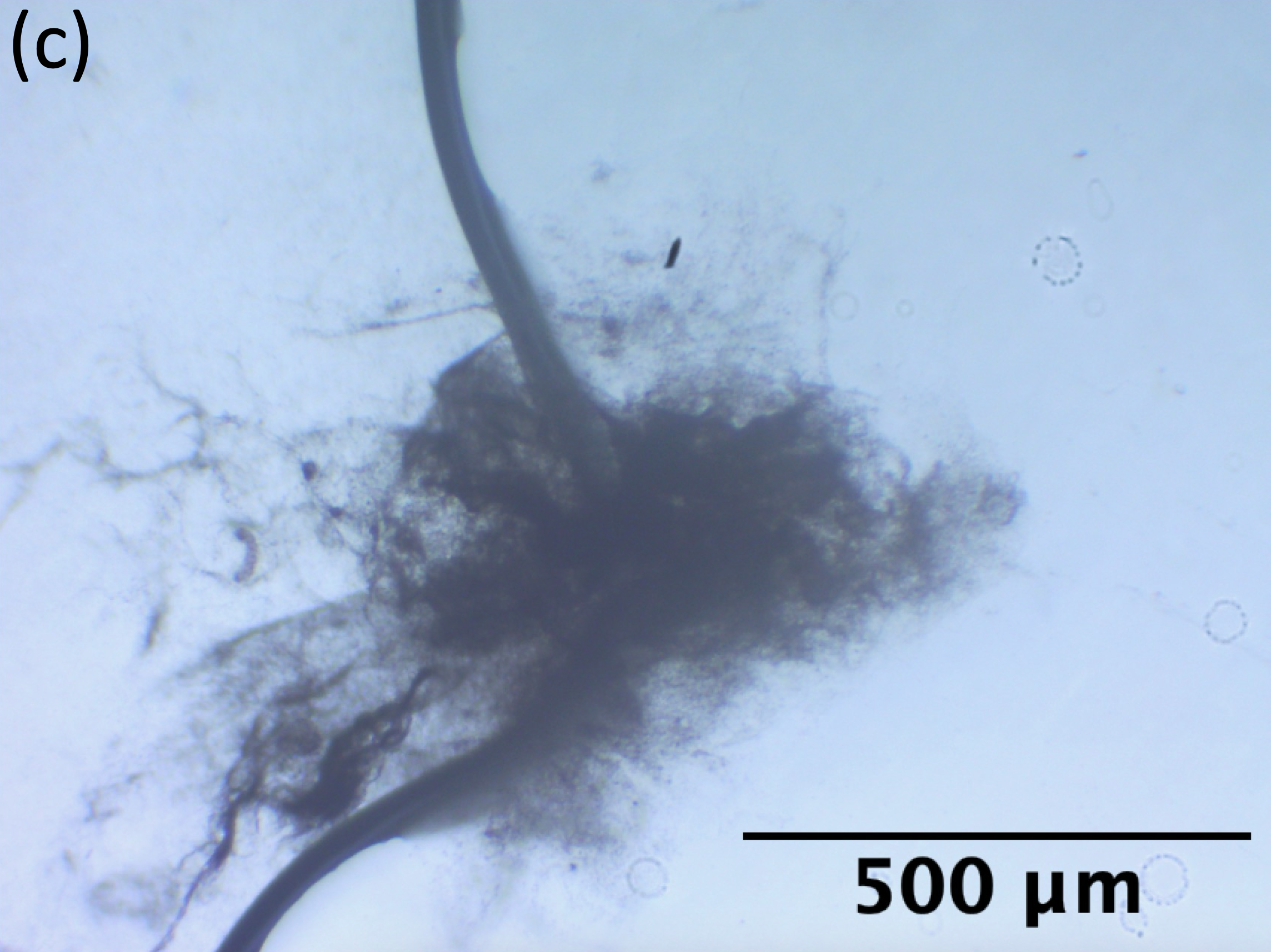}
    \includegraphics[width=0.3\columnwidth]{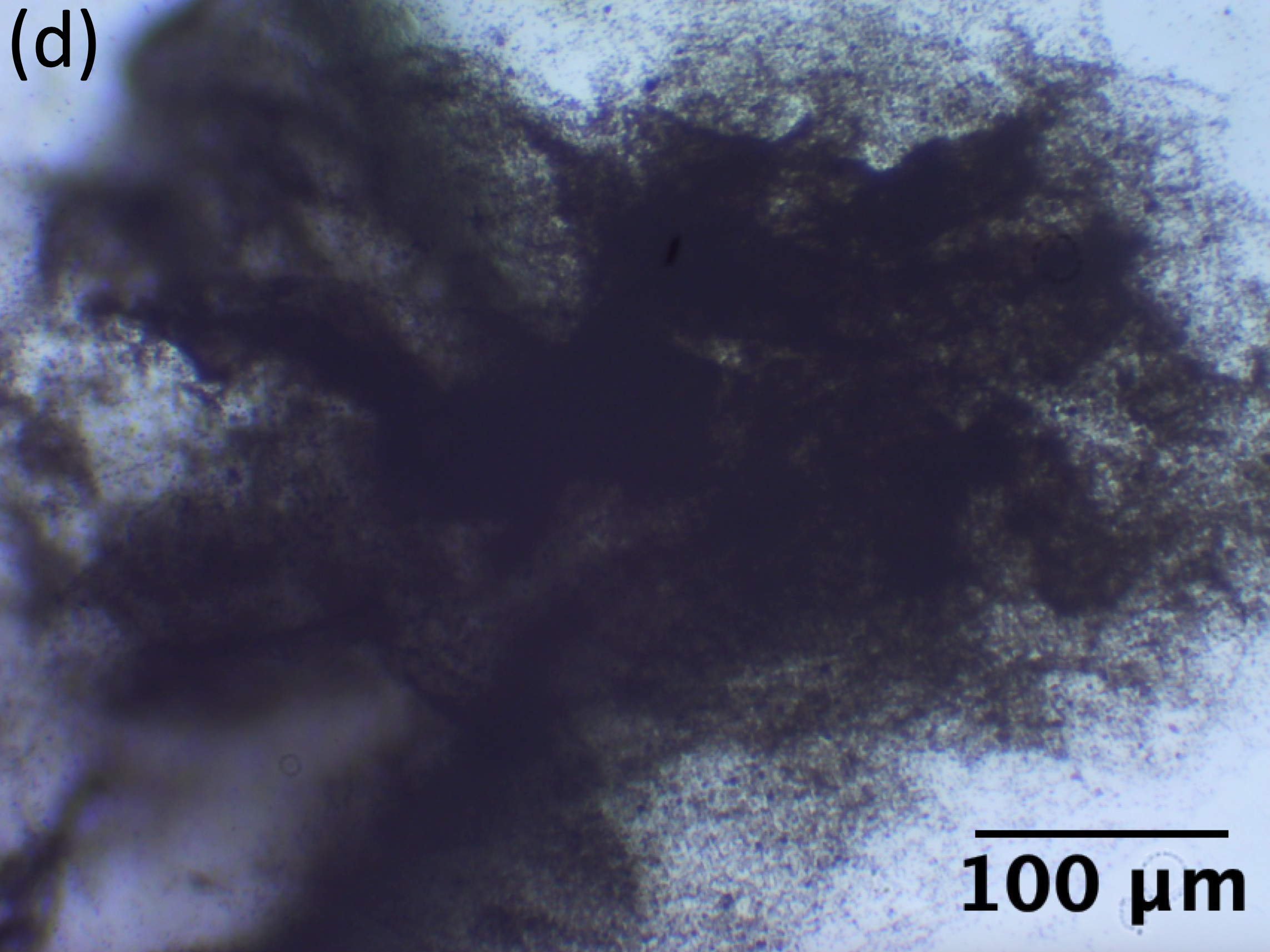}
    \includegraphics[width=0.3\columnwidth]{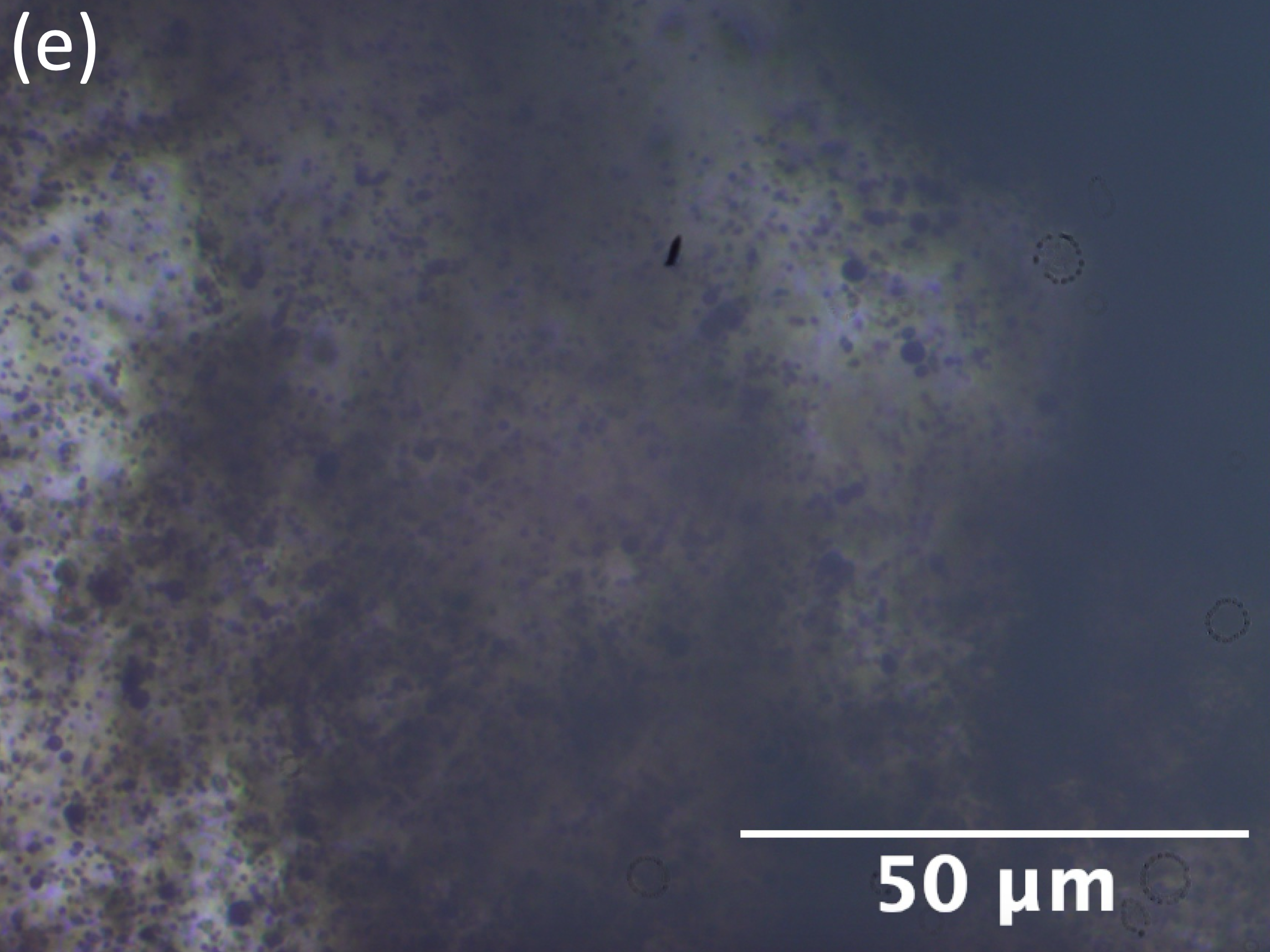}
    \caption{Optical microscope images of melanosome organelles observed under different magnifications. (a) Planar sample of unaggregated melanosomes showing uniform distribution observed under 40x magnification. (b) Volumetric distribution of unaggregated melanosomes, where overlapping regions result in a blurred appearance due to depth-of-field limitations observed under 40x magnification. (c) 4x magnification reveals large clusters of aggregated melanosomes. (d) 10x magnification provides a closer view of the aggregation patterns. (e) 40x magnification shows detailed structures within the aggregates, indicating possible interactions or binding between organelles.}
    \label{fig:pic1}
\end{figure}

During the microscopy analysis, we made several interesting visual observations regarding the spatial arrangement and behavior of melanosomes. In some cases, we noticed that melanosomes tended to cluster together (Fig. \ref{fig:pic1} (c, d, e), forming tight groups or aggregates. This clustering behavior may indicate interactions between melanosomes or local changes in the sample environment.

At other times, melanosomes were more uniformly distributed across the sample (Fig. \ref{fig:pic1} (a, b)), appearing as individual particles separated from one another. This homogeneous separation suggests a more stable dispersion, potentially influenced by the preparation process or environmental factors. When the concentration of melanosomes was high, multiple layers formed, and focusing on one layer resulted in melanosomes in other layers appearing blurred.

Additionally, in some observations, melanosomes appeared to be connected by a viscous liquid (Fig. \ref{fig:pic1}), forming a network-like structure. This could be due to residual components in the sample, such as proteins or lipids, that form a gel-like matrix surrounding the melanosomes. These variations in melanosome behavior under the microscope provide valuable context for understanding their dynamic nature and interactions during laser exposure.


\section*{Obtaining decay constants through reconvolution fit}

The Savitzky-Golay smoothing filter \cite{savitzky1964smoothing} is a widely used technique for smoothing a set of data points, effectively preserving the shape and features of the signal, such as peaks and valleys, better than other averaging filters. This filter works by fitting successive subsets of adjacent data points with a low-degree polynomial through the method of linear least squares. Savitzky-Golay filter was applied to all collected Signal and IRF before analyzing.

\subsection*{Reconvolution Fitting Process}

The reconvolution fitting process is a crucial step in analyzing TCSPC data. The goal is to model the observed fluorescence decay curve by deconvolving the IRF from the measured signal, allowing accurate estimation of the decay parameters of the sample \cite{vevcevr1993reconvolution}.

Reconvolution fitting is essential for accurately extracting decay parameters from TCSPC data. In this approach, the measured signal \(S(t)\) is modeled as the convolution of the sample’s actual fluorescence decay \(F(t)\) and the instrument response function \(R(t)\), i.e.\ \(\displaystyle S(t)=\int_{0}^{\infty} F(t')\,R(t-t')\,dt'\). The decay is typically represented as a sum of exponentials, \(\displaystyle F(t)=\sum_{i=1}^{n}A_{i} e^{-t/\tau_{i}}\), and the IRF \cite{counting2009time} is aligned using a non-integer shift if necessary. To find the best-fit parameters (decay constants \(\tau_{i}\), amplitudes \(A_{i}\), and IRF shift \(\Delta t\)), we employ a global optimization (e.g., Differential Evolution \cite{price2013differential}), followed by Non-Negative Least Squares (NNLS). This process ensures robust fitting of the observed TCSPC data and allows for precise determination of fluorescence lifetimes. 

\subsection*{Fit Validation with 4-DASPI}
\begin{figure*}[h!]
    \centering
    \includegraphics[width=0.75\textwidth]{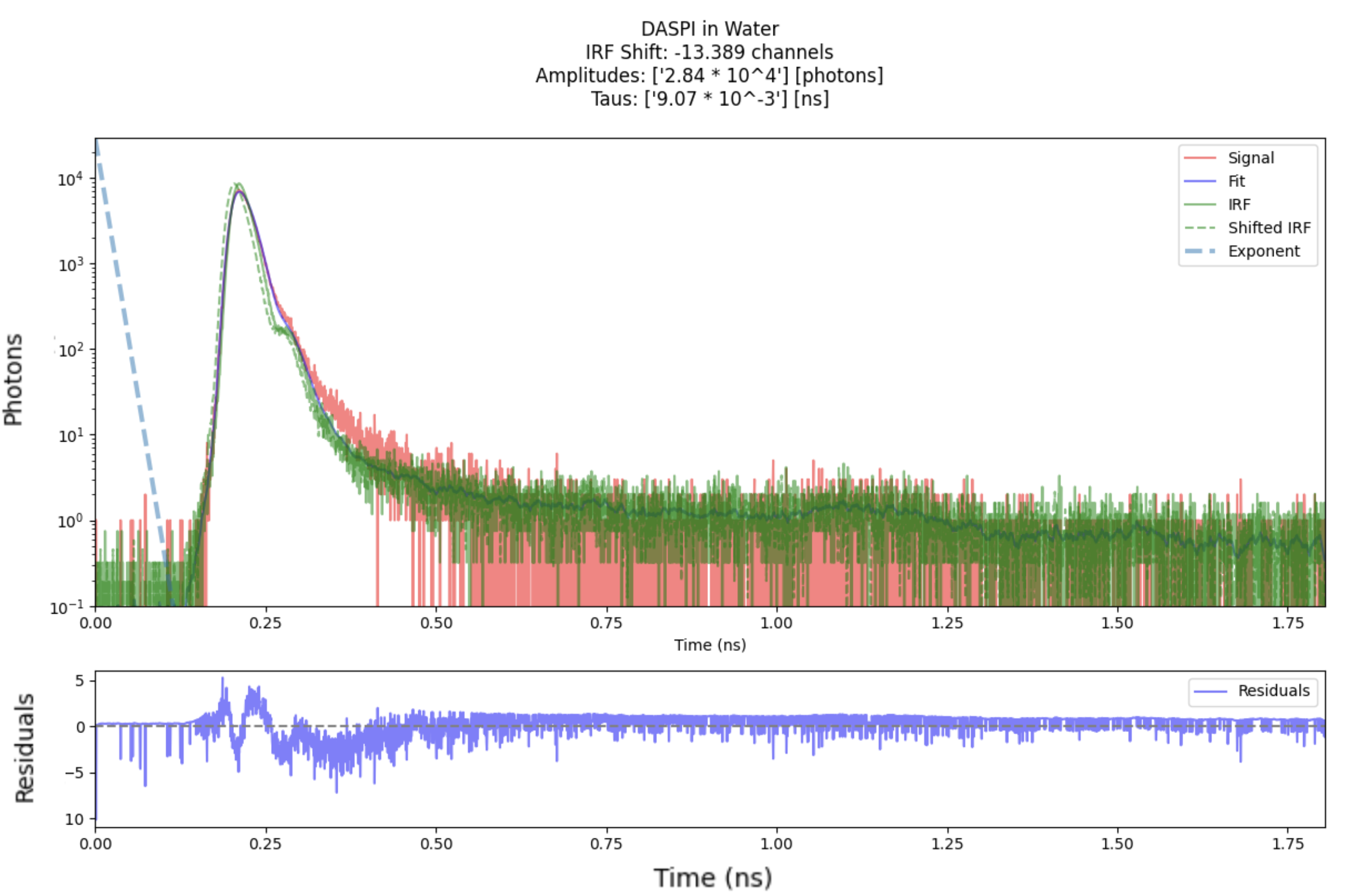}
    \caption{Reconvolution fit of the TCSPC data for 4-DASPI in water. The red line represents the signal, the green line is the IRF, and the blue line is the reconvolution fit.}
    \label{fig:DASPI_Reconvolution}
\end{figure*}
To validate the reconvolution fitting process, we used 4-DASPI dissolved in water as a standard reference. 4-DASPI is known for its short fluorescence lifetime in water at around 10 $ps$ \cite{kim1999excited, elbert2021dendron}. 

We performed a TCSPC measurement on a 25 $\mu L$ sample of 4-DASPI diluted in water, under the same experimental conditions as for the melanosome measurements. The fluorescence decay curve was fitted using the same reconvolution procedure, where the IRF was measured in real-time.

This validation step provides confidence in the fitting procedure and ensures that the decay times for melanosome fluorescence lifetimes can be accurately retrieved.



\section*{Data Analysis}

We continuously imaged melanosomes for 60 minutes at 1-minute intervals under $0.35\ \mathrm{mW}$ laser exposure. We specifically focused on melanosomes that were immobilized between the concave slide and coverslip, minimizing movement due to Brownian motion \cite{uhlenbeck1930theory}. Throughout the imaging process, changes in the fluorescence signal were observed, including variations in intensity and peak shape, which reflect dynamic shifts in both fluorescence intensity and lifetime.

To reduce the impact of Brownian motion on fluorescence intensity measurements, we allowed the sample to partially dry before imaging, reducing mobility. Additionally, the excitation volume was relatively large compared to individual melanosomes, mitigating signal fluctuations caused by minor displacements. We selected the most immobilized melanosomes for analysis and confirmed their position remained stable within the field of view throughout the experiment.

\begin{figure*}[h!]
    \centering
    \includegraphics[width=0.75\textwidth]{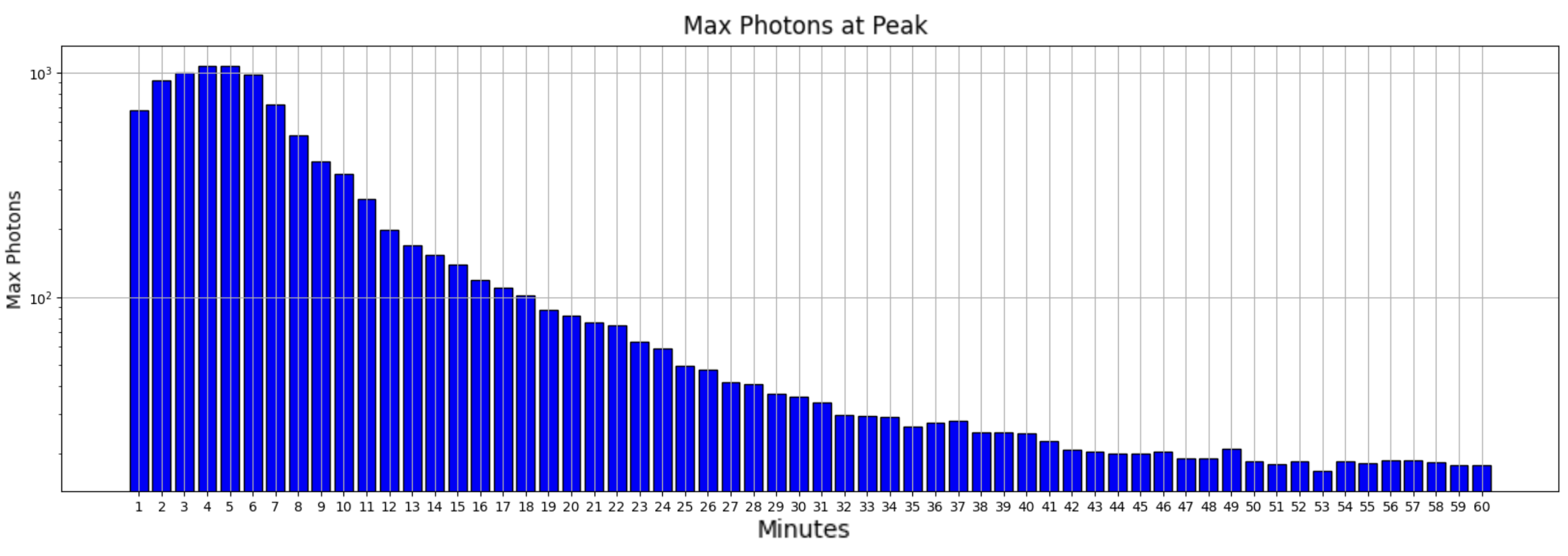} \\ 
    \includegraphics[width=0.75\textwidth]{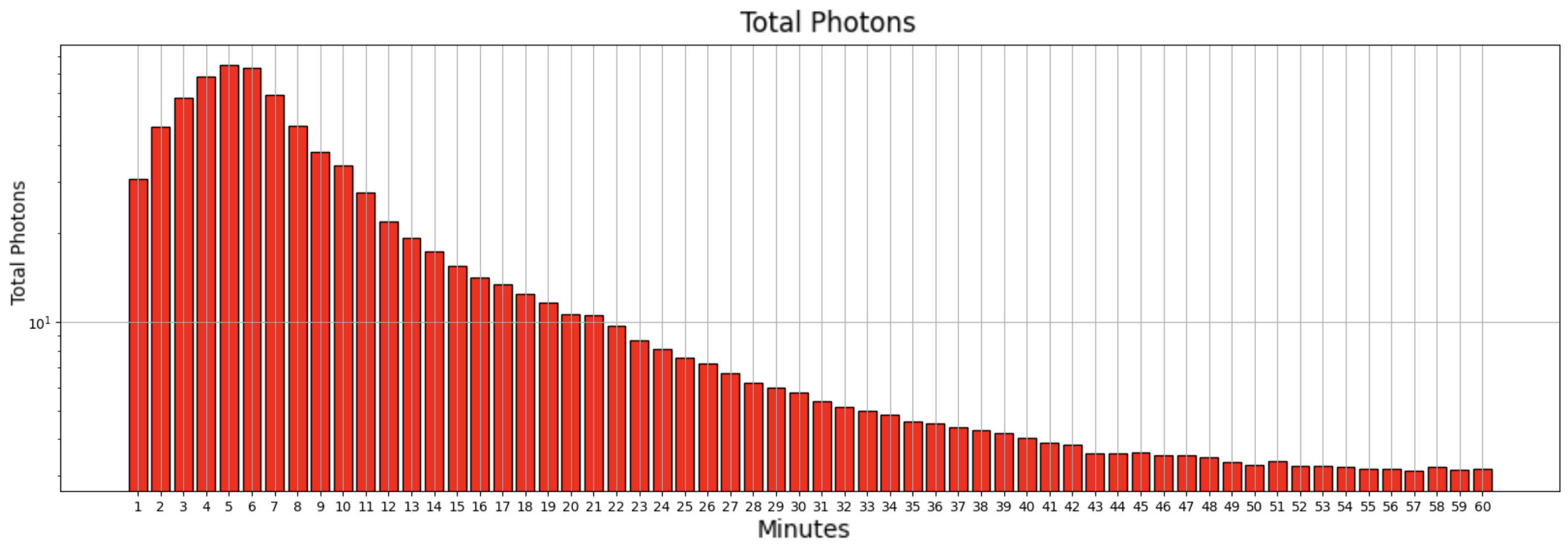} \\ 
    \includegraphics[width=0.75\textwidth]{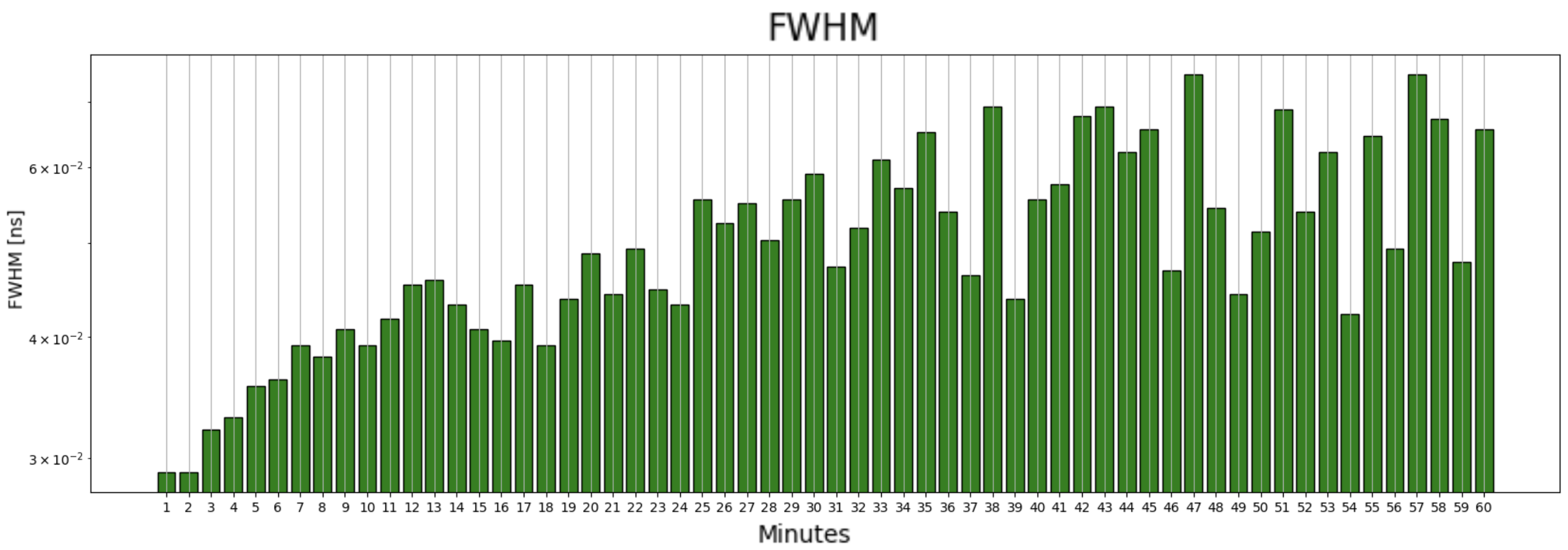}
    \caption{Initial rise followed by a decline. Top: Peak photon counts per minute. Middle: Total photons detected per minute. Bottom: FWHM of spectra per minute.}
    \label{fig:prong_2}
\end{figure*}

During continuous imaging of melanosomes over 60 minutes, we identified multiple patterns in their fluorescence response, including random variations likely caused by Brownian motion \cite{uhlenbeck1930theory}, gradual or monotonic changes in fluorescence parameters, and complex combinations of multiple processes. However, in this study, we focus on the most prominent pattern observed. 
This rise-and-fall pattern, depicted in Fig. \ref{fig:prong_2}, can be attributed to several causes. Laser-induced activation and degradation may explain the phenomenon, with initial fluorescence increase due to enhanced molecular excitation, and subsequent decline due to degradation or molecular rearrangement. Photobleaching from continuous laser exposure might cause gradual loss of fluorescence intensity. Additionally, thermal effects could initially boost fluorescence, later decreasing due to molecular denaturation or damage. The implications of this pattern highlight a combination of activation and degradation processes, illustrating the dynamic nature of melanosome response to laser exposure.

It is worth noting that even when fluorescence variation might be the same, total intensity might still be different \cite{Fazel2022}. We can observe this discrepancy mainly because of Brownian motion. Despite the fact that we tried to aim at the most 'still' melanosomes, we did not completely eliminate this effect.

For each recording that we have, we performed reconvolution fit to extract decay constants. 

In Fig. \ref{fig:prong_2}, we can observe the evolution of key parameters during imaging. The total number of detected photons, peak photon counts, and the FWHM of the fluorescence spectra are plotted for each minute of imaging. While the total and peak photon counts exhibit an initial rise followed by a decline, the FWHM continues to increase over time. This suggests that distinct photophysical processes influence spectral broadening and intensity changes in melanosomes during prolonged laser exposure.

\begin{figure*}[h!]
    \centering
    \includegraphics[width=0.75\textwidth]{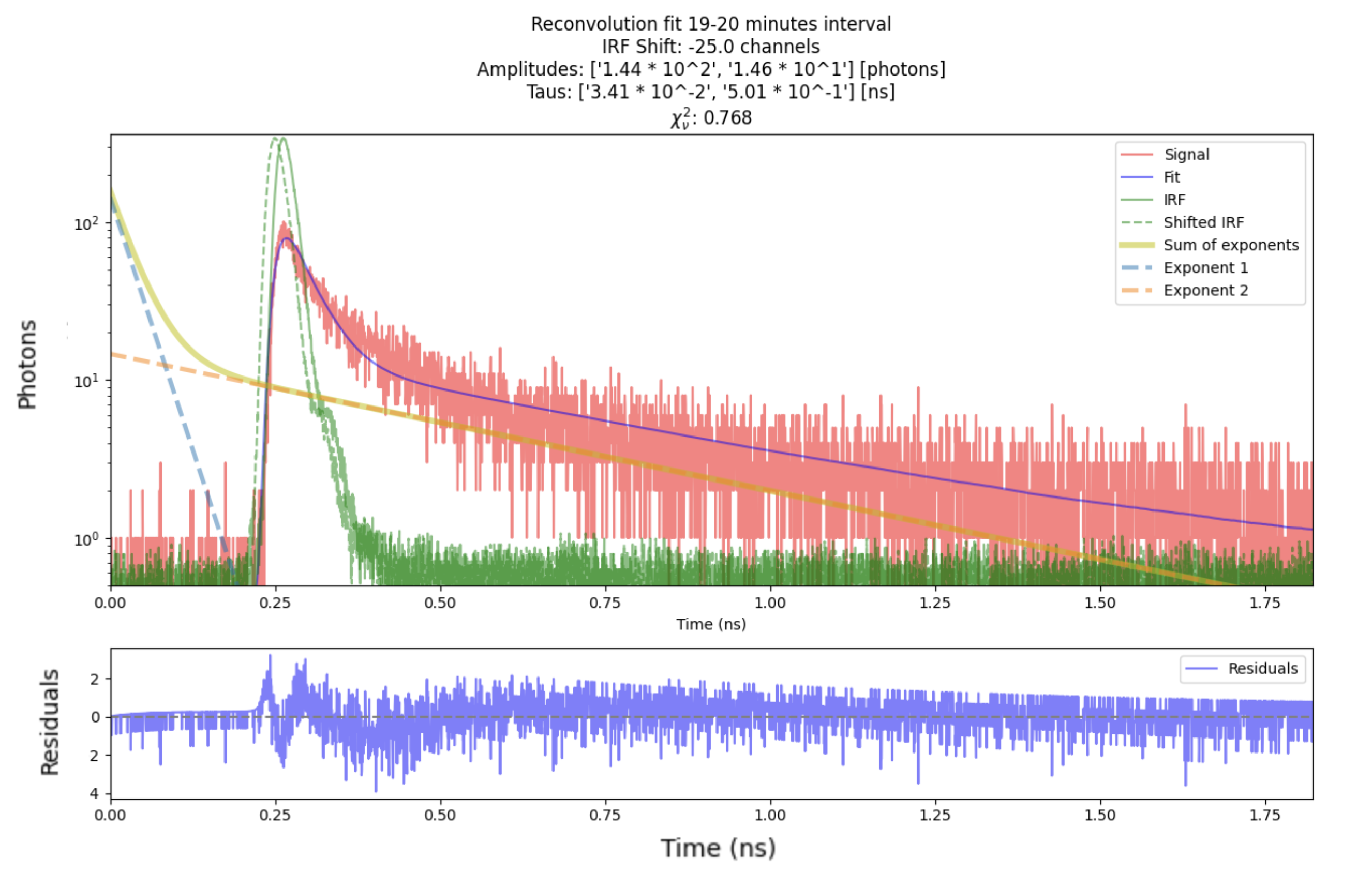}
    \caption{Reconvolution fit of the fluorescence decay data for melanosomes after 20 minutes of laser exposure, corresponding to the data in Fig. \ref{fig:prong_2}. The measured decay curve (dots) and the fitted model (solid line) are shown, illustrating the extraction of decay constants $\tau_1$ and $\tau_2$. Residuals indicate the goodness of fit achieved by the reconvolution method.}
    \label{fig:reconvolution_1_20}
\end{figure*}

To extract the fluorescence lifetimes, we performed a reconvolution fitting procedure on the fluorescence decay curves. A sample fit is presented in Fig. \ref{fig:reconvolution_1_20}, showing the quality of the reconvolution for a 19-20 minute interval. The fitting process allowed us to extract two lifetime components, $\tau_1$ and $\tau_2$, which describe the fast and slow decay processes, respectively.

\begin{figure}[h!]
    \centering
    \includegraphics[width=0.75\columnwidth]{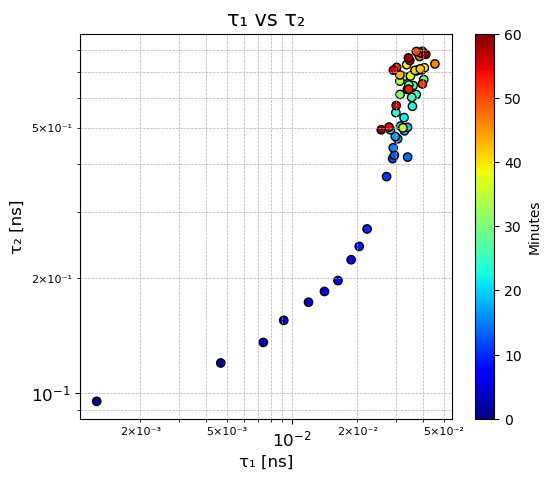}
    \caption{Evolution of the fluorescence lifetime components $\tau_1$ (fast decay) and $\tau_2$ (slow decay) over 60 minutes of continuous laser exposure, corresponding to the data in Fig. \ref{fig:prong_2}. Both lifetime components increase steadily over time, suggesting structural or chemical changes in melanosomes due to laser-induced effects.}
    \label{fig:taus_1}
\end{figure}

As shown in Fig. \ref{fig:taus_1}, both lifetime components, $\tau_1$ and $\tau_2$, increased steadily over the course of the 60 minutes of laser exposure. Notably, the variability in lifetime estimates becomes more pronounced for lifetimes greater than 0.03 ns. This increased scattering for longer lifetimes is likely due to the difficulty in accurately estimating slow decay components with fast FLIM acquisition. The estimated uncertainty in extracted lifetime values is approximately ±10\% for a single measurement.

To support our interpretation of fluorescence dynamics, the full set of decay constants ($\tau_1$, $\tau_1$) and their corresponding amplitudes ($A_1$, $A_2$) extracted from reconvolution fits are provided in \textit{Supplementary Table 1}. This table includes numerical values for all 60 time points corresponding to Fig.  \ref{fig:taus_1}, allowing detailed inspection of trends and fitting robustness.

It should be noted that the $\chi^2$ values obtained from the reconvolution fitting are consistently below 1 (e.g., 0.768 in \ref{fig:reconvolution_1_20}). This is primarily due to the application of the Savitzky–Golay filter, which effectively smooths the data and reduces noise, thereby lowering the $\chi^2$ value. We emphasize that this low $\chi^2$ reflects the high quality of our fits, which are consistent across all measurements. Potential sources of deviation may stem from differences in the probing geometry; the IRF is measured using a reflective geometry on a glass slide, whereas the melanosome sample, being volumetric, is probed using the full numerical aperture of the objective. Such geometric differences can lead to subtle variations in the extracted lifetimes.

\begin{figure*}[h!]
  \centering
  \includegraphics[width=0.8\textwidth]{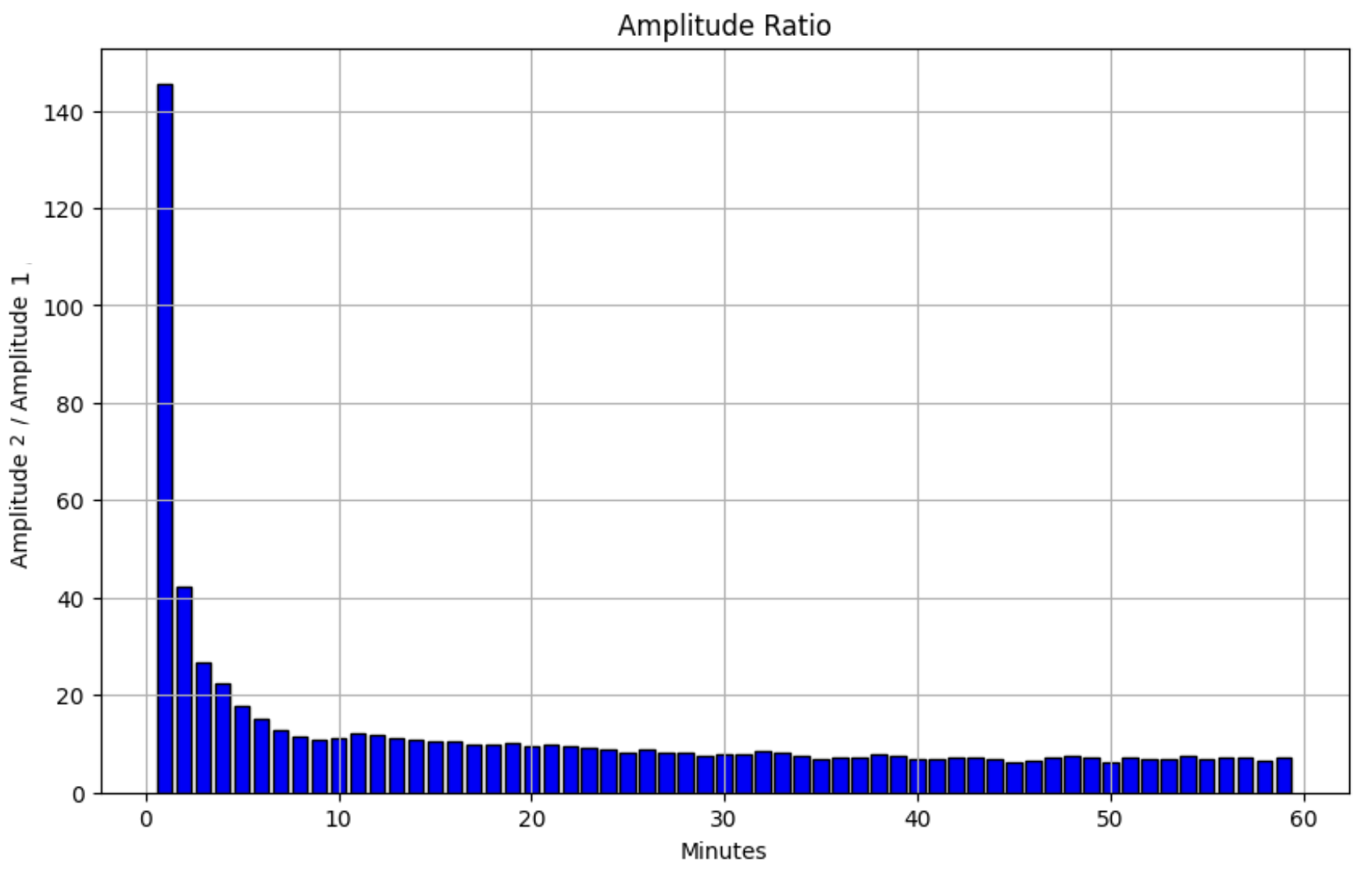}
  \caption{Temporal evolution of the amplitude ratio (A$_2$/A$_1$) between the fast and slow decay components over 60 minutes of laser exposure, indicating changes in the relative contributions of the two processes. The initial rapid decrease suggests a shift in fluorescence dynamics.}
  \label{fig:ampl_1}
\end{figure*}

The ratio of the amplitudes corresponding to the two lifetime components is plotted in Fig. \ref{fig:ampl_1}. Initially, the amplitude ratio decreases rapidly within the first 10 minutes of imaging, after which it stabilizes and remains relatively constant for the remainder of the experiment.

After performing reconvolution fitting for 40 FLIM measurements over 60 minutes, we extracted the decay constants ($\tau_1$ and $\tau_2$) for each melanosome sample. The extracted decay parameters revealed a notable correlation between the two time constants, which is visualized in Fig. \ref{fig:taus_1}.

The extracted fluorescence decay components, $\tau_1$ and $\tau_2$, represent two prominent exponential decay processes in melanosomes. However, we currently do not have direct evidence linking these lifetimes to specific degradation pathways such as oxidation or polymer rearrangement. Instead, the presence of two distinct lifetime components suggests heterogeneity in the fluorescent states or molecular environments within melanosomes. Further studies will be needed to directly correlate these lifetimes with known biochemical transformations in melanin.

To further investigate the behavior of melanosomes, we applied a robust fitting model using a second-order polynomial with Huber loss minimization. This allowed us to more accurately characterize the relationship between the fast ($\tau_1$) and slow ($\tau_2$) decay processes.

The analysis of the amplitude ratios provided additional insights into the fluorescence behavior of melanosomes over time. The largest amplitude ratios were observed in regions with the longest decay constants, suggesting that prolonged laser exposure leads to more prominent slow-decay components in the melanosome fluorescence.

The observed trend further emphasizes that melanosomes undergo cumulative photodamage during prolonged laser exposure, which may compromise their photoprotective function. This is especially significant for clinical applications involving laser-based therapies, where precise control of exposure parameters is crucial to minimize unintended tissue damage. Future studies could benefit from expanding the range of laser power levels to identify specific thresholds for melanosome degradation. Furthermore, incorporating multi-wavelength FLIM imaging could differentiate between eumelanin- and pheomelanin-dominant melanosomes, offering deeper insights into pigment-specific degradation dynamics.
\section*{Discussion}

Photophysics and photochemistry of melanins have been the subject of extensive research; however, relatively limited efforts have been devoted to studies of melanosomes due to their structural and chemical complexity. In this report, we employ fast FLIM \cite{Hirvonen2020, Sun2024, Tan2023} to provide novel insights into structural changes in melanosomes under laser exposure. FLIM is uniquely capable of sensing the local nanoenvironment and is relatively unaffected by photobleaching. Both short- and long-lifetime components of fluorescence emission were recorded, reflecting the complexity of molecular processes. The complex changes of fluorescence decays were observed over minutes of laser irradiation, signifying the initiation of multiple processes such as melanin oxidation, polymer reorganization, and local structure changes that are triggered by the laser excitation.

The initial increase observed in fluorescence intensity \cite{Pena2023} indicates that laser-induced activation of selected melanin substructures could transiently increase fluorescence output. As exposure continues, partial photobleaching probably takes place, whereas local heating might result in self-annealing of local structural defects, which initially promoted a short-lived fluorescence.

While photobleaching could contribute to the observed changes in fluorescence lifetime, the steady increase in both lifetime components over time suggests an underlying structural or chemical transformation rather than simple fluorophore depletion. At longer radiation exposures, gradual and steady changes in fluorescence lifetimes are likely the result of progressive oxidative modifications of melanin.

In evaluating the complex nature of FLIM dynamics upon laser irradiation, we developed a novel reconvolution software tool to facilitate robust analysis of FLIM data. By automating the process of deconvolving the instrument response function from measured fluorescence decay curves, this tool streamlines the extraction of fluorescence lifetime components across diverse experimental conditions. This software is openly accessible on GitHub \cite{github}, enabling wider adoption of advanced reconvolution-based methods for both research and clinical applications.

These observations are supported by amplitude ratios (A$_2$/A$_1$) and the evolution of fast ($\tau_1$) and slow ($\tau_2$) decay constants. In many samples, an early shift toward higher A$_1$ suggests that fast-decay pathways momentarily dominate during initial laser exposure. As exposure lengthens, a higher fraction of slow-decay components emerges, suggesting the accumulation of photoproducts, cross-linked regions, or other reorganized melanin motifs. This interpretation aligns with prior work \cite{Herrera-Ochoa2024, Ashworth2023} indicating that longer-lived fluorescent states often arise from either trapped excited states or more rigid environments. Notably, temperature-induced changes and oxidative stress can stabilize these states, increasing fluorescence lifetimes beyond initial baseline levels.

Some limitations of our experimental design include variability associated with Brownian motion, which impeded fluorescence signal stability despite attempts to immobilize the melanosomes. Additionally, the current experimental setup was limited to a single laser power and wavelength, constraining the scope of our analysis.

To overcome these limitations, optical trapping or microfluidic devices could limit sample movement and increase data consistency, while multispectral FLIM studies, varying laser power and wavelengths, would differentiate photophysical and photochemical effects more clearly. Multispectral FLIM could also distinguish between eumelanin and pheomelanin, providing further insights into pigment-specific behavior under laser exposure \cite{Meyer2019, dake2018timedomain}.
FLIM can be adapted for in vivo imaging of melanosomes in tissue by leveraging its ability to measure the fluorescence decay times of naturally occurring fluorescent molecules within the skin. This technique, when combined with multiphoton laser tomography (MPT), allows for non-invasive imaging with subcellular resolution, which is crucial for detailed examination of melanosomes. In the context of melanoma diagnostics, FLIM can differentiate between melanoma and other skin lesions by identifying atypical short lifetime cells and architectural disorder, which are characteristic of melanoma, as opposed to the regular histoarchitecture seen in nevi. This capability enhances diagnostic accuracy, achieving high sensitivity and specificity, and could potentially reduce unnecessary surgical procedures by providing a more precise non-invasive diagnostic tool \cite{Seidenari2013Multiphoton}. 

\section*{Conclusion}

This study successfully demonstrated the application of FLIM to investigate the degradation dynamics of melanosomes under laser exposure. By analyzing fluorescence lifetime components, both short- and long-lived, we identified distinct photophysical changes that correlate with structural and biochemical alterations in melanosomes. The observed trends, such as the increase in fluorescence decay times and variations in amplitude ratios, strongly indicate oxidative modifications and thermal denaturation as the primary contributors to laser-induced degradation. These findings align with previous studies highlighting the utility of FLIM for capturing dynamic cellular processes \cite{ma2020highspeed} and monitoring molecular changes in real-time \cite{ryu2018realtime}.

The heterogeneity in fluorescence patterns reflects the complex interplay of molecular excitation, photobleaching, and chemical reactions, emphasizing the need for advanced imaging techniques to resolve these dynamics. The insights provided by this study are crucial for optimizing laser parameters in clinical applications, particularly in dermatology and ophthalmology, to minimize tissue damage while achieving therapeutic goals \cite{benard2021optimization}.

Notable novelties introduced by this study include the direct visualization and quantification of melanosome degradation in real-time, enhancing our understanding of photodamage at the organelle level. Additionally, we detected a picosecond-scale decay component, offering new insights into ultrafast photophysical dynamics associated with melanin-containing structures.

In summary, fast FLIM is minimally invasive and provides dynamic information about structural changes in melanosomes on a time scale from minutes to hours. This study bridges gaps in understanding photodamage mechanisms and paves the way for safer and more effective laser-based therapies. The findings presented lay the foundation for future explorations of FLIM in clinical and biomedical research, enabling real-time monitoring of molecular interactions and dynamic processes in living tissues.

\section*{Acknowledgments}
This work was partially supported by the Air Force of Scientific Research (grants FA9550-20-1-0366 and FA9550-23-1-0599), the National Institutes of Health (grants R01GM127696, R21GM142107, 1R21CA269099), and the U.S. Food and Drug Administration (80ARC023CA002E)\\
\begin{shaded}
\noindent\textsf{\textbf{Keywords:} \keywords} 
\end{shaded}

\bibliographystyle{IEEEtran}
\bibliography{reference}

\begin{thebibliography}{10}
\providecommand{\url}[1]{#1}
\csname url@samestyle\endcsname
\providecommand{\newblock}{\relax}
\providecommand{\bibinfo}[2]{#2}
\providecommand{\BIBentrySTDinterwordspacing}{\spaceskip=0pt\relax}
\providecommand{\BIBentryALTinterwordstretchfactor}{4}
\providecommand{\BIBentryALTinterwordspacing}{\spaceskip=\fontdimen2\font plus
\BIBentryALTinterwordstretchfactor\fontdimen3\font minus \fontdimen4\font\relax}
\providecommand{\BIBforeignlanguage}[2]{{%
\expandafter\ifx\csname l@#1\endcsname\relax
\typeout{** WARNING: IEEEtran.bst: No hyphenation pattern has been}%
\typeout{** loaded for the language `#1'. Using the pattern for}%
\typeout{** the default language instead.}%
\else
\language=\csname l@#1\endcsname
\fi
#2}}
\providecommand{\BIBdecl}{\relax}
\BIBdecl

\bibitem{day2005imaging}
R.~N. Day and F.~Schaufele, ``Imaging molecular interactions in living cells,'' \emph{Molecular endocrinology}, vol.~19, no.~7, pp. 1675--1686, 2005.

\bibitem{gomperts2009signal}
B.~D. Gomperts, P.~E. Tatham \emph{et~al.}, \emph{Signal transduction}.\hskip 1em plus 0.5em minus 0.4em\relax Academic Press, 2009.

\bibitem{kang2008molecular}
J.~H. Kang and J.-K. Chung, ``Molecular-genetic imaging based on reporter gene expression,'' \emph{Journal of Nuclear Medicine}, vol.~49, no. Suppl 2, pp. 164S--179S, 2008.

\bibitem{liu2012imaging}
J.~Liu, Y.~Xu, D.~Stoleru, and A.~Salic, ``Imaging protein synthesis in cells and tissues with an alkyne analog of puromycin,'' \emph{Proceedings of the National Academy of Sciences}, vol. 109, no.~2, pp. 413--418, 2012.

\bibitem{house2009tracking}
D.~House, M.~L. Walker, Z.~Wu, J.~Y. Wong, and M.~Betke, ``Tracking of cell populations to understand their spatio-temporal behavior in response to physical stimuli,'' in \emph{2009 IEEE Computer Society Conference on Computer Vision and Pattern Recognition Workshops}.\hskip 1em plus 0.5em minus 0.4em\relax IEEE, 2009, pp. 186--193.

\bibitem{meng2016subcellular}
Z.~Meng, S.~C. Bustamante~Lopez, K.~E. Meissner, and V.~V. Yakovlev, ``Subcellular measurements of mechanical and chemical properties using dual raman-brillouin microspectroscopy,'' \emph{Journal of Biophotonics}, vol.~9, no.~3, pp. 201--207, 2016.

\bibitem{liu2015imaging}
Z.~Liu, L.~D. Lavis, and E.~Betzig, ``Imaging live-cell dynamics and structure at the single-molecule level,'' \emph{Molecular cell}, vol.~58, no.~4, pp. 644--659, 2015.

\bibitem{2025CheburkanovBrillouinSPIE}
\BIBentryALTinterwordspacing
V.~Cheburkanov, M.~Kizilov, S.~Jung, M.~Y. Berezin, and V.~V. Yakovlev, ``\BIBforeignlanguage{English}{Non-invasive remote assessment of tissue fibrogenesis using brillouin microscopy},'' in \emph{\BIBforeignlanguage{English}{Optical Elastography and Tissue Biomechanics XII}}, K.~V. Larin and G.~Scarcelli, Eds., vol. 13321, International Society for Optics and Photonics.\hskip 1em plus 0.5em minus 0.4em\relax SPIE, 2025, p. 133210H. [Online]. Available: \url{https://doi.org/10.1117/12.3058161}
\BIBentrySTDinterwordspacing

\bibitem{2025CheburkanovGliaSPIE}
\BIBentryALTinterwordspacing
V.~Cheburkanov, M.~Kizilov, S.~Jung, K.~I.~O. Sandoval, S.~Raghavan, and V.~Yakovlev, ``\BIBforeignlanguage{English}{Noninvasive investigation of enteric glia culture viscoelastic properties},'' in \emph{\BIBforeignlanguage{English}{Label-free Biomedical Imaging and Sensing (LBIS) 2025}}, N.~T. Shaked and O.~Hayden, Eds., vol. 13331, International Society for Optics and Photonics.\hskip 1em plus 0.5em minus 0.4em\relax SPIE, January 26 2025, p. 1333106. [Online]. Available: \url{https://doi.org/10.1117/12.3044067}
\BIBentrySTDinterwordspacing

\bibitem{burke2005mosaicism}
J.~M. Burke and L.~M. Hjelmeland, ``Mosaicism of the retinal pigment epithelium: seeing the small picture,'' \emph{Molecular interventions}, vol.~5, no.~4, p. 241, 2005.

\bibitem{prota2012melanins}
G.~Prota, \emph{Melanins and melanogenesis}.\hskip 1em plus 0.5em minus 0.4em\relax Academic Press, 2012.

\bibitem{miltenberger2002molecular}
R.~J. Miltenberger, K.~Wakamatsu, S.~Ito, R.~P. Woychik, L.~B. Russell, and E.~J. Michaud, ``Molecular and phenotypic analysis of 25 recessive, homozygous-viable alleles at the mouse agouti locus,'' \emph{Genetics}, vol. 160, no.~2, pp. 659--674, 2002.

\bibitem{sarna1992new}
T.~Sarna, ``New trends in photobiology: properties and function of the ocular melanin—a photobiophysical view,'' \emph{Journal of Photochemistry and Photobiology B: Biology}, vol.~12, no.~3, pp. 215--258, 1992.

\bibitem{hill1997melanin}
H.~Z. Hill, W.~Li, P.~Xin, and D.~L. Mitchell, ``Melanin: a two edged sword?'' \emph{Pigment cell research}, vol.~10, no.~3, pp. 158--161, 1997.

\bibitem{Alam2022}
S.~R. Alam, H.~Wallrabe, K.~G. Christopher, K.~Siller, and A.~Periasamy, ``Characterization of phototoxic effects in multiphoton flim,'' \emph{SPIE Proceedings}, vol. 11965, pp. 119\,650B -- 119\,650B--7, 2022.

\bibitem{zareba2006effects}
M.~Zareba, G.~Szewczyk, T.~Sarna, L.~Hong, J.~D. Simon, M.~M. Henry, and J.~M. Burke, ``Effects of photodegradation on the physical and antioxidant properties of melanosomes isolated from retinal pigment epithelium,'' \emph{Photochemistry and photobiology}, vol.~82, no.~4, pp. 1024--1029, 2006.

\bibitem{Dontsov2023understanding}
A.~E. Dontsov, M.~A. Yakovleva, A.~A. Vasin, A.~A. Gulin, A.~V. Aybush, V.~A. Nadtochenko, and M.~A. Ostrovsky, ``Understanding the mechanism of light-induced age-related decrease in melanin concentration in retinal pigment epithelium cells,'' \emph{International Journal of Molecular Sciences}, vol.~24, no.~17, p. 13099, 2023.

\bibitem{Jimbow1982Characterization}
K.~Jimbow, M.~Jimbow, and M.~Chiba, ``Characterization of structural properties for morphological differentiation of melanosomes: Ii. electron microscopic and sds-page comparison of melanosomal matrix proteins in b16 and harding passey melanomas,'' \emph{Journal of Investigative Dermatology}, vol.~78, pp. 76--81, 1982.

\bibitem{Lea1976The}
P.~Lea, H.~Haberman, A.~Pawlowski, and I.~Menon, ``The use of freeze-fracture and negative staining techniques to study the ultrastructure of melanosomes isolated from b16 melanoma.'' \emph{Journal of anatomy}, vol. 121 Pt 1, pp. 1--5, 1976.

\bibitem{ruedas2015flim}
M.~J. Ruedas-Rama, J.~M. Alvarez-Pez, L.~Crovetto, J.~M. Paredes, and A.~Orte, ``Flim strategies for intracellular sensing: Fluorescence lifetime imaging as a tool to quantify analytes of interest,'' \emph{Advanced Photon Counting: Applications, Methods, Instrumentation}, pp. 191--223, 2015.

\bibitem{vasanthakumari2022discrimination}
P.~Vasanthakumari, R.~A. Romano, R.~G. Rosa, A.~G. Salvio, V.~Yakovlev, C.~Kurachi, J.~M. Hirshburg, and J.~A. Jo, ``Discrimination of cancerous from benign pigmented skin lesions based on multispectral autofluorescence lifetime imaging dermoscopy and machine learning,'' \emph{Journal of biomedical optics}, vol.~27, no.~6, pp. 066\,002--066\,002, 2022.

\bibitem{vasanthakumari2024pixel}
------, ``Pixel-level classification of pigmented skin cancer lesions using multispectral autofluorescence lifetime dermoscopy imaging,'' \emph{Biomedical Optics Express}, vol.~15, no.~8, pp. 4557--4583, 2024.

\bibitem{2025KizilovColonAPS}
M.~Kizilov, V.~Cheburkanov, and V.~V. Yakovlev, ``Differentiation of wild-type and crispr-modified colon cancer cells using brillouin microscopy,'' in \emph{APS March Meeting Abstracts}, vol. 2025, 2025, pp. OD01--012, arXiv preprint arXiv:2507.05329.

\bibitem{duran2021machine}
E.~Duran-Sierra, S.~Cheng, R.~Cuenca, B.~Ahmed, J.~Ji, V.~V. Yakovlev, M.~Martinez, M.~Al-Khalil, H.~Al-Enazi, Y.-S.~L. Cheng \emph{et~al.}, ``Machine-learning assisted discrimination of precancerous and cancerous from healthy oral tissue based on multispectral autofluorescence lifetime imaging endoscopy,'' \emph{Cancers}, vol.~13, no.~19, p. 4751, 2021.

\bibitem{schmidt2016temperature}
M.~S. Schmidt, P.~Kennedy, G.~Noojin, R.~Thomas, and B.~Rockwell, ``Temperature dependence of nanosecond laser pulse thresholds of melanosome and microsphere microcavitation,'' \emph{Journal of Biomedical Optics}, vol.~21, 2016.

\bibitem{yi2018degraded}
W.~Yi, M.~Su, Y.~Shi, S.~Jiang, S.-z. Xu, and T.~Lei, ``Degraded melanocores are incompetent to protect epidermal keratinocytes against uv damage,'' \emph{Cell Cycle}, vol.~17, pp. 844--857, 2018.

\bibitem{hu2020heat}
W.~Hu, N.~Mi, Y.~Xu, G.~Zhao, and W.~Gu, ``42 °c heat stress pretreatment protects human melanocytes against 308-nm laser-induced dna damage in vitro,'' \emph{Lasers in Medical Science}, vol.~35, pp. 1801--1809, 2020.

\bibitem{luecking2020capabilities}
M.~Luecking, R.~Brinkmann, S.~Ramos, W.~Stork, and N.~Heussner, ``Capabilities and limitations of a new thermal finite volume model for the evaluation of laser-induced thermo-mechanical retinal damage,'' \emph{Computers in Biology and Medicine}, vol. 122, p. 103835, 2020.

\bibitem{alghamdi2016ultrastructural}
K.~Alghamdi, A.~Kumar, A.~Al-ghamdi, A.~Al-rikabi, M.~Mubarek, and A.~Ashour, ``Ultra-structural effects of different low-level lasers on normal cultured human melanocytes: an in vitro comparative study,'' \emph{Lasers in Medical Science}, vol.~31, pp. 1819--1825, 2016.

\bibitem{saha2011raman}
A.~Saha, R.~Arora, V.~V. Yakovlev, and J.~M. Burke, ``Raman microspectroscopy of melanosomes: the effect of long term light irradiation,'' \emph{Journal of Biophotonics}, vol.~4, no. 11-12, pp. 805--813, 2011.

\bibitem{yakovlev2008real}
V.~V. Yakovlev, R.~J. Thomas, G.~Noojin, and M.~Denton, ``Real-time monitoring of chemical and structural changes induced by light irradiation of cells and tissues,'' in \emph{Imaging, Manipulation, and Analysis of Biomolecules, Cells, and Tissues VI}, vol. 6859.\hskip 1em plus 0.5em minus 0.4em\relax SPIE, 2008, pp. 90--97.

\bibitem{suhling2015fluorescence}
K.~Suhling, L.~M. Hirvonen, J.~A. Levitt, P.-H. Chung, C.~Tregidgo, A.~Le~Marois, D.~A. Rusakov, K.~Zheng, S.~Ameer-Beg, S.~Poland \emph{et~al.}, ``Fluorescence lifetime imaging (flim): Basic concepts and some recent developments,'' \emph{Medical photonics}, vol.~27, pp. 3--40, 2015.

\bibitem{2025CheburkanovHemoglobinSPIE}
\BIBentryALTinterwordspacing
V.~Cheburkanov, M.~Kizilov, and V.~Yakovlev, ``\BIBforeignlanguage{English}{Fluorescence lifetime spectroscopy of hemoglobin},'' in \emph{\BIBforeignlanguage{English}{Optical Diagnostics and Sensing XXV: Toward Point-of-Care Diagnostics}}, G.~L. Cot{\'e} and J.~S. Baba, Eds., vol. 13316, International Society for Optics and Photonics.\hskip 1em plus 0.5em minus 0.4em\relax SPIE, 2025, p. 133160K. [Online]. Available: \url{https://doi.org/10.1117/12.3044088}
\BIBentrySTDinterwordspacing

\bibitem{2025BerezinFLIMSPIE}
\BIBentryALTinterwordspacing
H.~Biswas, R.~Tang, M.~Michie, M.~Kizilov, V.~Cheburkanov, V.~V. Yakovlev, and M.~Y. Berezin, ``\BIBforeignlanguage{English}{Non-fitting algorithms for fluorescence lifetime imaging},'' in \emph{\BIBforeignlanguage{English}{Reporters, Contrast Agents, and Molecular Probes for Biomedical Applications XVI}}, M.~Y. Berezin and R.~Raghavachari, Eds., vol. PC13339, International Society for Optics and Photonics.\hskip 1em plus 0.5em minus 0.4em\relax SPIE, 2025, p. PC1333902. [Online]. Available: \url{https://doi.org/10.1117/12.3053224}
\BIBentrySTDinterwordspacing

\bibitem{2025KizilovMelanomaSPIE}
M.~Kizilov, V.~Cheburkanov, S.~Jung, and V.~V. Yakovlev, ``Viscoelastic characterization of melanoma cells using brillouin spectroscopy,'' in \emph{APS March Meeting Abstracts}, vol. 2025, 2025, pp. OD01--011, arXiv preprint arXiv:2507.05186.

\bibitem{yakovlev2018biochemical}
V.~Yakovlev, \emph{Biochemical applications of nonlinear optical spectroscopy}.\hskip 1em plus 0.5em minus 0.4em\relax CRC Press, 2018.

\bibitem{Kizilov2025}
\BIBentryALTinterwordspacing
M.~Kizilov, S.~Jung, V.~Cheburkanov, and V.~Yakovlev, ``\BIBforeignlanguage{English}{Fluorescence lifetime imaging and signal reconvolution for characterizing laser-induced melanosome degradation},'' in \emph{\BIBforeignlanguage{English}{Multimodal Biomedical Imaging XX}}, X.~Intes, M.~Ochoa, and M.~A. Yaseen, Eds., vol. 13309, International Society for Optics and Photonics.\hskip 1em plus 0.5em minus 0.4em\relax SPIE, 2025, p. 133090L. [Online]. Available: \url{https://doi.org/10.1117/12.3048829}
\BIBentrySTDinterwordspacing

\bibitem{Shimojo2024}
Y.~Shimojo, T.~Nishimura, D.~Tsuruta, T.~Ozawa, H.~H.~L. Chan, and T.~Kono, ``Wavelength‐dependent threshold fluences for melanosome disruption to evaluate the treatment of pigmented lesions with 532‐, 730‐, 755‐, 785‐, and 1064‐nm picosecond lasers,'' \emph{Lasers in Surgery and Medicine}, vol.~56, pp. 404--418, 2024.

\bibitem{Kauvar2012}
A.~Kauvar, ``The evolution of melasma therapy: targeting melanosomes using low-fluence q-switched neodymium-doped yttrium aluminium garnet lasers,'' \emph{Seminars in Cutaneous Medicine and Surgery}, vol.~31, pp. 126--132, 2012.

\bibitem{jhraman}
J.~Harrington, V.~Cheburkanov, M.~Kizilov, I.~Kulagin, G.~Petrov, and V.~V. Yakovlev, ``Highly-sensitive, low-cost duv resonant raman microspectroscopy system,'' \emph{Chemistry-Methods}, 2025.

\bibitem{mykraman}
\BIBentryALTinterwordspacing
M.~Kizilov, V.~Cheburkanov, J.~Harrington, and V.~Yakovlev, ``\BIBforeignlanguage{English}{Advanced preprocessing and analysis techniques for enhanced raman spectroscopy data interpretation},'' in \emph{\BIBforeignlanguage{English}{Optical Biopsy XXIII: Toward Real-Time Spectroscopic Imaging and Diagnosis}}, R.~R. Alfano, A.~B. Seddon, L.~Shi, and B.~Wu, Eds., vol. 13311, International Society for Optics and Photonics.\hskip 1em plus 0.5em minus 0.4em\relax SPIE, 2025, p. 133110F. [Online]. Available: \url{https://doi.org/10.1117/12.3048830}
\BIBentrySTDinterwordspacing

\bibitem{Cubeddu1990}
R.~Cubeddu, F.~Docchio, R.~Ramponi, and M.~Boulton, ``Time-resolved fluorescence spectroscopy of the retinal pigment epithelium: age-related studies,'' \emph{IEEE Journal of Quantum Electronics}, vol.~26, pp. 2218--2225, 1990.

\bibitem{2025HarringtonDUVSPIE}
\BIBentryALTinterwordspacing
J.~T. Harrington, V.~Cheburkanov, M.~Kizilov, I.~Kulagin, G.~Petrov, and V.~V. Yakovlev, ``\BIBforeignlanguage{English}{Deep ultraviolet resonant raman (duvrr) spectroscopy for spectroscopic evaluation and disinfection of food and agricultural samples},'' in \emph{\BIBforeignlanguage{English}{Photonic Technologies in Plant and Agricultural Science II}}, vol. 13357, International Society for Optics and Photonics.\hskip 1em plus 0.5em minus 0.4em\relax SPIE, 2025, p. 1335702. [Online]. Available: \url{https://doi.org/10.1117/12.3042310}
\BIBentrySTDinterwordspacing

\bibitem{meleppat2021vivo}
R.~K. Meleppat, K.~E. Ronning, S.~J. Karlen, M.~E. Burns, E.~N. Pugh~Jr, and R.~J. Zawadzki, ``In vivo multimodal retinal imaging of disease-related pigmentary changes in retinal pigment epithelium,'' \emph{Scientific reports}, vol.~11, no.~1, p. 16252, 2021.

\bibitem{dontsov2024retinal}
A.~Dontsov and M.~Ostrovsky, ``Retinal pigment epithelium pigment granules: Norms, age relations and pathology,'' \emph{International Journal of Molecular Sciences}, vol.~25, no.~7, p. 3609, 2024.

\bibitem{Seidenari2013}
S.~Seidenari, F.~Arginelli, C.~Dunsby, P.~French, K.~König, C.~Magnoni, C.~Talbot, and G.~Ponti, ``Multiphoton laser tomography and fluorescence lifetime imaging of melanoma: Morphologic features and quantitative data for sensitive and specific non-invasive diagnostics,'' \emph{PLoS ONE}, vol.~8, 2013.

\bibitem{Arginelli2013}
F.~Arginelli, M.~Manfredini, S.~Bassoli, C.~Dunsby, P.~French, K.~König, C.~Magnoni, G.~Ponti, C.~Talbot, and S.~Seidenari, ``High resolution diagnosis of common nevi by multiphoton laser tomography and fluorescence lifetime imaging,'' \emph{Skin Research and Technology}, vol.~19, 2013.

\bibitem{Fink2016}
C.~Fink, M.~Hofmann, A.~Jagoda, I.~Spaenkuch, A.~Forschner, I.~Tampouri, D.~Lomberg, D.~Leupold, C.~Garbe, and H.~Haenssle, ``Study protocol for a prospective, non-controlled, multicentre clinical study to evaluate the diagnostic accuracy of a stepwise two-photon excited melanin fluorescence in pigmented lesions suspicious for melanoma (flimma study),'' \emph{BMJ Open}, vol.~6, 2016.

\bibitem{cone2015measuring}
M.~T. Cone, J.~D. Mason, E.~Figueroa, B.~H. Hokr, J.~N. Bixler, C.~C. Castellanos, G.~D. Noojin, J.~C. Wigle, B.~A. Rockwell, V.~V. Yakovlev \emph{et~al.}, ``Measuring the absorption coefficient of biological materials using integrating cavity ring-down spectroscopy,'' \emph{Optica}, vol.~2, no.~2, pp. 162--168, 2015.

\bibitem{furtjes2023intraoperative}
G.~F{\"u}rtjes, D.~Reinecke, N.~von Spreckelsen, A.-K. Mei{\ss}ner, D.~Rue{\ss}, M.~Timmer, C.~Freudiger, A.~Ion-Margineanu, F.~Khalid, K.~Watrinet \emph{et~al.}, ``Intraoperative microscopic autofluorescence detection and characterization in brain tumors using stimulated raman histology and two-photon fluorescence,'' \emph{Frontiers in Oncology}, vol.~13, p. 1146031, 2023.

\bibitem{dontsov1999retinal}
A.~E. Dontsov, R.~D. Glickman, and M.~A. Ostrovsky, ``Retinal pigment epithelium pigment granules stimulate the photo-oxidation of unsaturated fatty acids,'' \emph{Free Radical Biology and Medicine}, vol.~26, no. 11-12, pp. 1436--1446, 1999.

\bibitem{denton2013hyperthermia}
M.~L. Denton, G.~D. Noojin, M.~S. Foltz, V.~V. Yakovlev, L.~E. Estlack, R.~J. Thomas, and B.~A. Rockwell, ``Hyperthermia sensitizes pigmented cells to laser damage without changing threshold damage temperature,'' \emph{Journal of Biomedical Optics}, vol.~18, no.~11, pp. 110\,501--110\,501, 2013.

\bibitem{roegener2004pump}
J.~Roegener, R.~Brinkmann, and C.~P. Lin, ``Pump-probe detection of laser-induced microbubble formation in retinal pigment epithelium cells,'' \emph{Journal of Biomedical Optics}, vol.~9, no.~2, pp. 367--371, 2004.

\bibitem{savitzky1964smoothing}
A.~Savitzky and M.~J. Golay, ``Smoothing and differentiation of data by simplified least squares procedures.'' \emph{Analytical chemistry}, vol.~36, no.~8, pp. 1627--1639, 1964.

\bibitem{vevcevr1993reconvolution}
J.~Ve{\v{c}}e{\v{r}}, A.~Kowalczyk, L.~Davenport, and R.~Dale, ``Reconvolution analysis in time-resolved fluorescence experiments—an alternative approach: Reference-to-excitation-to-fluorescence reconvolution,'' \emph{Review of scientific instruments}, vol.~64, no.~12, pp. 3413--3424, 1993.

\bibitem{counting2009time}
W.~Becker, A.~Bergmann, M.~Hink, K.~K{\"o}nig, K.~Benndorf, and C.~Biskup, ``Fluorescence lifetime imaging by time-correlated single-photon counting,'' \emph{Microscopy research and technique}, vol.~63, no.~1, pp. 58--66, 2004.

\bibitem{price2013differential}
K.~V. Price, ``Differential evolution,'' in \emph{Handbook of optimization: From classical to modern approach}.\hskip 1em plus 0.5em minus 0.4em\relax Springer, 2013, pp. 187--214.

\bibitem{kim1999excited}
J.~Kim and M.~Lee, ``Excited-state photophysics and dynamics of a hemicyanine dye in aot reverse micelles,'' \emph{The Journal of Physical Chemistry A}, vol. 103, no.~18, pp. 3378--3382, 1999.

\bibitem{elbert2021dendron}
K.~C. Elbert, ``Dendron design, synthesis, and application in nanocrystal assembly,'' Ph.D. dissertation, University of Pennsylvania, 2021.

\bibitem{uhlenbeck1930theory}
G.~E. Uhlenbeck and L.~S. Ornstein, ``On the theory of the brownian motion,'' \emph{Physical review}, vol.~36, no.~5, p. 823, 1930.

\bibitem{Fazel2022}
M.~Fazel, S.~Jazani, L.~Scipioni, A.~Vallmitjana, E.~Gratton, M.~Digman, and S.~Pressé, ``High resolution fluorescence lifetime maps from minimal photon counts,'' \emph{ACS Photonics}, vol.~9, pp. 1015 -- 1025, 2022.

\bibitem{Hirvonen2020}
L.~Hirvonen and K.~Suhling, ``Fast timing techniques in flim applications,'' \emph{Frontiers in Physics}, vol.~8, 2020.

\bibitem{Sun2024}
M.-J. Sun, Y.-C. Zhang, F.~Lin, S.~Wang, L.~Liu, and J.~Qu, ``Rapid fluorescence lifetime imaging microscopy via few-photon imaging,'' \emph{APL Photonics}, 2024.

\bibitem{Tan2023}
K.~K.~D. Tan, M.~A. Tsuchida, J.~Chacko, N.~A. Gahm, and K.~Eliceiri, ``Real-time open-source flim analysis,'' \emph{Frontiers in Bioinformatics}, vol.~3, 2023.

\bibitem{Pena2023}
A.~Pena, S.~Ito, T.~Bornschlögl, S.~Brizion, K.~Wakamatsu, and S.~Del~Bino, ``Multiphoton flim analyses of native and uva-modified synthetic melanins,'' \emph{International Journal of Molecular Sciences}, vol.~24, 2023.

\bibitem{github}
M.~Kizilov, ``Flim reconvolution toolkit,'' \url{https://github.com/mkizilov/ReconFit}.

\bibitem{Herrera-Ochoa2024}
D.~Herrera-Ochoa, I.~Llano, C.~Ripoll, P.~Cybulski, M.~Kreuzer, S.~Rocha, E.~M. García-Frutos, I.~Bravo, and A.~Garzón‐Ruiz, ``Protein aggregation monitoring in cells under oxidative stress: a novel fluorescent probe based on a 7-azaindole-bodipy derivative,'' \emph{Journal of materials chemistry. B}, 2024.

\bibitem{Ashworth2023}
E.~K. Ashworth, M.-H. Kao, C.~S. Anstöter, G.~Riesco-Llach, L.~Blancafort, K.~M. Solntsev, S.~Meech, J.~Verlet, and J.~Bull, ``Alkylated green fluorescent protein chromophores: dynamics in the gas phase and in aqueous solution,'' \emph{Physical chemistry chemical physics : PCCP}, 2023.

\bibitem{Meyer2019}
T.~Meyer, H.~Bae, S.~Hasse, J.~Winter, T.~Woedtke, M.~Schmitt, K.~Weltmann, and J.~Popp, ``Multimodal nonlinear microscopy for therapy monitoring of cold atmospheric plasma treatment,'' \emph{Micromachines}, vol.~10, 2019.

\bibitem{dake2018timedomain}
F.~Dake and Y.~Taki, ``Time-domain fluorescence lifetime imaging by nonlinear fluorescence microscopy constructed of a pump-probe setup with two-wavelength laser pulses,'' \emph{Applied Optics}, vol.~57, pp. 757--762, 2018.

\bibitem{Seidenari2013Multiphoton}
S.~Seidenari, F.~Arginelli, C.~Dunsby, P.~French, K.~König, C.~Magnoni, C.~Talbot, and G.~Ponti, ``Multiphoton laser tomography and fluorescence lifetime imaging of melanoma: Morphologic features and quantitative data for sensitive and specific non-invasive diagnostics,'' \emph{PLoS ONE}, vol.~8, 2013.

\bibitem{ma2020highspeed}
Y.~Ma, Y.~Lee, C.~Best-Popescu, and L.~Gao, ``High-speed compressed-sensing fluorescence lifetime imaging microscopy of live cells,'' \emph{Proceedings of the National Academy of Sciences}, vol. 118, 2020.

\bibitem{ryu2018realtime}
J.~Ryu, U.~Kang, and J.~e.~a. Kim, ``Real-time visualization of two-photon fluorescence lifetime imaging microscopy using a wavelength-tunable femtosecond pulsed laser,'' \emph{Biomedical Optics Express}, vol.~9, pp. 3449--3463, 2018.

\bibitem{benard2021optimization}
M.~e.~a. Bénard, ``Optimization of advanced live-cell imaging through red/near-infrared dye labeling and fluorescence lifetime-based strategies,'' \emph{International Journal of Molecular Sciences}, vol.~22, 2021.

\end{thebibliography}

\clearpage
\end{document}